\documentstyle[preprint,aps,axodraw,epsf]{revtex}
\newcommand{\postscript}[2] {\setlength{\epsfxsize}{#2\hsize}
\centerline{\epsfbox{#1}}}
\begin{document}
\title{Light-Front Bethe-Salpeter Equation}
\author{J.H.O. Sales$^a$, T.Frederico$^a$, B.V.Carlson$^a$, P.U. Sauer$^b$}
\address{
$^a$ Dep.de F\'\i sica, Instituto Tecnol\'ogico de Aeron\'autica, \\
Centro T\'ecnico Aeroespacial, \\
12.228-900 S\~ao Jos\'e dos Campos, S\~ao Paulo, Brazil.\\
$^b$Institute for Theoretical Physics, University Hannover \\
D-30167 Hannover, Germany}
\date{\today }
\maketitle
\begin{abstract}
A three-dimensional reduction of the two-particle Bethe-Salpeter equation is
proposed. The proposed reduction is in the framework of light-front
dynamics. It yields auxiliary quantities for the transition matrix and the
bound state. The arising effective interaction can be perturbatively
expanded according to the number of particles exchanged at a given
light-front time. An example suggests that the convergence of the
expansion is rapid. This result is particular for light-front dynamics. The
covariant results of the Bethe-Salpeter equation can be recovered from the
corresponding auxiliary three-dimensional ones. The technical procedure is
developed for a two-boson case; the idea for an extension to fermions is
given. The technical procedure appears quite practicable, possibly allowing one
to go beyond the ladder approximation for the solution of the Bethe-Salpeter
equation. The relation between the three-dimensional light-front reduction
of the field-theoretic Bethe-Salpeter equation and a corresponding
quantum-mechanical description is discussed.
\end{abstract}

\pacs{12.39.Ki,14.40.Cs,13.40.Gp}

\section{Introduction}

In relativistic field theory the Bethe-Salpeter equation (BSE) \cite{bs}
describes two-particle systems in interaction. The inhomogeneous BSE 
\begin{equation}
T=V+VG_0T  \label{1.1}
\end{equation}
yields the transiton matrix $T$ of two-particle scattering. In Eq.(\ref{1.1}%
) $G_0$ is the free resolvent which propagates two non-interacting particles, 
i.e.,
\begin{equation}
G_0=\frac i{\widehat{k}_1^2-m_1^2+io }\frac i{\widehat{k}_2^2-m_2^2+io },
\label{1.4a}
\end{equation}
$\widehat{k}_i^\mu $ denoting the momentum operator of particle $i$ with
mass $m_i$, the hat on the variable emphasizing its operator character. The
driving term $V$ stands for the complete interaction, irreducible with
respect to two-particle propagation; it also includes self-energy
corrections, i.e., it may contain disconnected pieces. If the dynamics
allows for a two-particle bound state $\left| \Psi \right\rangle $ with
total four-momentum $K_B$ , $K_B^2=M_B^2$, the vertex $\left| \Gamma
\right\rangle $ at the bound-state pole is solution of the homogeneous BSE 
\begin{equation}
\left| \Gamma\right\rangle =VG_0\left| \Gamma\right\rangle  \label{1.2a}
\end{equation}
with the relation 
\begin{equation}
\left| \Psi \right\rangle =G_0\left| \Gamma\right\rangle \   \label{1.2b}
\end{equation}
to the bound state $\left| \Psi \right\rangle$. Eqs. (\ref{1.1}) and (\ref
{1.2a}) do not determine the bound state $\left| \Psi \right\rangle $ in
full; the normalization condition has to be added. The two-particle total
four-momentum $K$ is conserved, i.e., all operators ${\cal O}_\alpha $ 
of Eqs. (\ref{1.1}) and (\ref{1.2a}), ${\cal O}_\alpha =T,G_0$ 
or $V$, and the states 
$\left| \Psi \right\rangle $ and $\left| \Gamma \right\rangle $ carry a
four-dimensional $\delta -$function in momentum space, i.e., 
\begin{equation}
\left\langle K^{\prime }\right| {\cal O}_\alpha \left| K\right\rangle
=\delta (K^\prime-K){\cal O}_\alpha (K),
\end{equation}
\begin{equation}
\left\langle K^{\prime }\right| \left. \Psi \right\rangle =\delta (K^{\prime
}-K_B)\left| \Psi _B\right\rangle ,
\end{equation}
\begin{equation}
\left\langle K^{\prime }\right| \left. \Gamma\right\rangle =\delta
(K^{\prime }-K_B)\left| \Gamma_B\right\rangle ,
\end{equation}
the reduced quantities depending parametrically on $K$, even if not spelled
out explicitly for the states $\left| \Gamma_B\right\rangle$ and $\left|
\Psi _B\right\rangle $. The reduced states $\left| \Psi _B\right\rangle $
and $\left| \Gamma _B\right\rangle $ belong to and the operators ${\cal O}%
_\alpha (K)$ act in a Hilbert space characterized by a four-dimensional
momentum $k^\mu $ or coordinate $x^\mu $. They satisfy the Eqs. (\ref{1.1})
and (\ref{1.2a}) in a corresponding fashion.

The inhomogeneous and homogeneous BSEs (\ref{1.1}) and (\ref{1.2a}) are
general and exact formulations for the scattering amplitude and bound state.
However, for any realistic field theory solution  of
the BSE constitutes a difficult
calculational task which has not been tackled in full. In practical
calculations, the driving term $V(K)$ has to be truncated to low orders of
particle exchange. In Euclidean space, the fermion case has only been solved
in ladder approximation \cite{flei}, i.e., with single particle exchange for
the driving term while the boson case has only been solved 
in ladder and crossed ladder approximation \cite{tjon}.
However, the step from the Euclidean-space to Minkowski-space solutions
requires a complicated analytic continuation \cite{zui}. Direct solutions in
Minkowski space are just now becoming available \cite{will97}.

In the light of the great calculational difficulties, three-dimensional
reductions of the BSE are still of high physics interest. The conceptual
sacrifices generated by the reduction can possibly be outweighed by the
gain in technical ease: One hopes to be able to include
physical phenomena which the
four-dimensional BSE with a highly truncated interaction is unable to
account for. For example, the three-dimensional Gross 
approach \cite{gross} allows only
one particle to propagate off-mass-shell, but it appears to go  
beyond the ladder approximation of BSE by single particle exchange
and to
include crossed exchanges implicitly; it manfestly preserves covariance.
Other reduction schemes give up covariance, which then must be recovered
through complicated correction schemes. 
An equal-time projection scheme has also
been explored for the pion-nucleon system which fulfills requirements of
covariance and discrete Poincar\`e symmetries \cite{pasca}. The papers by
Fuda \cite{fuda} report on the comparision of one-meson exchange models in
ladder approximation on both light-front and instant-form dynamics,
without emphasis to the underlying field-theoretic framework.

The purpose of this paper is two-fold:

{\it i}) First, the paper attempts to find a three-dimensional equation for
auxiliary quantities from which the full covariant solution of the BSE in the
ladder or any other approximation can be obtained {\it with ease}. This is a
technical objective with solutions well-known in the framework of
instant-form dynamics. Here the advantages of light-front dynamics are to be
explored.

{\it ii}) Second, the paper tries to illuminate the connection to a
quantum-mechanical description of the two-particle system whose dynamic
input is related to the underlying field theory.

Sect.II motivates our novel choice for three-dimensional auxiliary
quantities from which the covariant solutions of the BSE are obtained. It
motivates light-front dynamics as our choice for a dynamical framework.
Sect.III gives our theoretical apparatus in full. Sect.IV tests the
potential of the method in the example of a two-boson bound state. We
perform numerical calculations for the two-boson bound state including up to
four-particle intermediate states in lowest order and compare to the
solutions of the four-dimensional BSE equation in the ladder approximation.
Sect.V sketches the generalization of our theoretical apparatus to fermions.
Sect.VI discusses the connection with light-front quantum mechanics. Our
conclusions are summarized in Sect.VII.

\section{Choice of two-particle auxiliary free 
resolvent $\widetilde{G}_0(k)$.}

It is well known, from the work of Ref.\cite{wolja}, that the transition
matrix $T(k)$ and the bound state $\left| \Psi _B\right\rangle $ of the
covariant BSE can be obtained with the help of a convenient auxiliary
resolvent $\widetilde{G}_0(K)$, still to be chosen. That is, we have 
\begin{eqnarray}
T(K) & = & W(K)+W(K)\widetilde{G}_0(K)T(K),  \label{2.1} \\
\left| \Gamma_B\right\rangle & = & W(K_B)\widetilde{G}_0(K_B)\left|
\Gamma_B\right\rangle ,  \label{2.2a} \\
\left| \Psi _B\right\rangle & = & G_0(K)\left| \Gamma_B\right\rangle \ ,
\label{2.2b}
\end{eqnarray}
provided the driving term $V(K)$ is changed to $W(K)$ according to 
\begin{equation}
W(K)=V(K)+V(K)[G_0(K)-\widetilde{G}_0(K)]W(K) \ .  \label{2.3a}
\end{equation}
Eqs. (\ref{2.2a}) and (\ref{2.2b}) do not determine the bound state $\left|
\Psi _B\right\rangle $ in full; the normalization condition 
\begin{eqnarray}
\lim_{K^2\rightarrow K^2_B} \left\langle \Psi_B \left| \frac{%
G_0(K)^{-1}-G_0(K_B)^{-1}}{K^2-K^2_B} -\frac{V(K)-V(K_B)}{K^2-K^2_B} \right|
\Psi _B\right\rangle=1  \label{3.6e}
\end{eqnarray}
has to be added. It involves the original driving term $V(K)$\cite{itz}. The
choice of $\widetilde{G}_0(K)$ is hoped to be sufficently 
clever that the integral
equation (\ref{2.3a}) does not have to be solved in full, but 
that a few terms of the infinite series 
\begin{eqnarray}
W(K) & = & V(K)\sum_{n=0}^\infty\left[ \left( G_0(K)- \widetilde G_0(K)
\right) V(K)\right] ^n  \nonumber \\
W(K) & = & V(K)+V(K)\left( G_0(K)-\widetilde G_0(K)\right) V(K)+...
\label{2.3b}
\end{eqnarray}
suffice. The auxiliary resolvent $\widetilde G_0(K)$ remains a
four-dimensional one, but its choice may sacrifice the covariance which the
resolvent $G_0(K)$ possesses.

The dynamics of the interacting two-particle system can be fully described
by its propagation between hyperplanes, the hyperplanes $x^0$= const. in
instant-form dynamics, the hyperplanes $x^{+}= x^0+x^3=$ const. in
light-front dynamics \cite{di49}. In contrast, the free resolvent of the BSE
depends on the individual times $x^0_i$ or on the individual light-front
times $x^+_i$.

The free resolvent in {\it instant-form} coordinates $k_i=(k^0_i,\vec k_i)$ 
\begin{eqnarray}
\left\langle x_1^{\prime 0}x_2^{\prime 0}\right| G_0\left|
x_1^0x_2^0\right\rangle &=&-\frac{1}{(2\pi)^2} \int dk_1^0dK^0\times 
\nonumber \\
&&\frac{e^{-ik_1^0\left( x_1^{\prime 0}-x_2^{\prime 0}-x_1^0+x_2^0\right) }}{%
\left((k_1^0)^2-\widehat{\vec k}^2-m_1^2+io\right)} \frac{e^{-iK^0\left(
x_2^{\prime 0}-x_2^0\right) }}{\left( (K^0-k_1^0)^2- (\vec K-\widehat{\vec k}%
_1)^2-m_2^2+io \right)}  \label{2.4}
\end{eqnarray}
$--$ in fact only its dependence on individual times $x^0_i$ is made
explicit $--$ reduces for propagation between the hyperplanes $x^0$ and $%
x^{\prime 0}$ to 
\begin{eqnarray}
\left\langle x^{\prime 0}x^{\prime 0}\right| G_0\left| x^0x^0\right\rangle
&=& \int \frac{dK^0}{2\pi} e^{-iK^0\left( x^{\prime 0}-x^{ 0}\right) } \int
dk^{\prime 0}_1 d k^0_1 \left\langle k_1^{\prime 0}\right|
G_0(K)\left|k^0_1\right\rangle \ ,  \label{2.5a} \\
&\equiv& \int \frac{dK^0}{2\pi} 
e^{-iK^0( x^{\prime 0}-x^{ 0}) } |_0 G_0(K) |_0 \ .  \label{2.5b}
\end{eqnarray}
In Eq.(\ref{2.5a}) the notation 
\begin{eqnarray}
\left\langle k_1^{\prime 0}\right| G_0(K)\left|k^0_1\right\rangle \ = -\frac1%
{2\pi} \frac{\delta \left(k^{\prime 0}_1-k^{ 0}_1\right) }{\left( (k_1^0)^2-%
\widehat{\vec k_1}^2-m_1^2+io\right) \left( (K^0-k_1^0)^2-(\vec K-\widehat{%
\vec k}_1)^2 -m^2_2+io\right) }  \label{2.6}
\end{eqnarray}
is introduced, as well as the abbreviation 
\begin{eqnarray}
|_0 G_0(K)|_0:&= & \int dk^{\prime 0}_1 dk^{ 0}_1 \left\langle k_1^{\prime
0}\right| G_0(K)\left|k^0_1\right\rangle  \label{2.7a} \\
&=&\frac{i}{2\widehat{k}^0_{1on} 2\widehat{k}^0_{2on}}\left( \frac1{%
\left(K^0-\widehat{k}_{1on}^0-\widehat{k}_{2on}^0+io \right)} -\frac1{%
\left(K^0+\widehat{k}_{1on}^0+\widehat{k}_{2on}^0-io \right)} \right)\ .
\label{2.7b}
\end{eqnarray}
The matrix element $\left\langle k_1^{\prime 0}\right|
G_0(K)\left|k^0_1\right\rangle $ of Eq.(\ref{2.6}), in which only the
dependence on the "dynamic" variable $k^0_1$ is made explicit remains
an operator with respect to the "kinematic" variables $\vec k_1$, the
operator character being carried by the operators $\widehat k^0_{ion}=\sqrt{%
\widehat{\vec k}_i^2+m_i^2}$. The basis states for these kinematic variables
are defined by $\langle \vec x_i \left| \vec k_i \right \rangle=
exp\left(\imath \vec k_i \cdot \vec x_i\right)$ and are eigenfunctions of
the momentum operator $\widehat{\vec k}$ and the free energy operator $%
\widehat k^0_{on} $ . The states $\left|\vec k \right\rangle$ form an
orthogonal and complete basis.

In Eq. (\ref{2.7a}), the vertical bar $|_0$ indicates that the dependence on 
$k^0_1$ is integrated out. The bar on the left of the resolvent represents
integration on $k^0_1$ in the bra-state, the bar on the right in the
ket-state; we shall encounter resolvents in which integration on $k^0_1$ is
performed only on one side, the bar $|_0$ being placed on that side alone. 
The resulting operator $|_0 G_0(K)|_0$ is three-dimensional 
and depends only on the
kinematic variables $\vec k_1$. It is a global propagator, since it
mediates between hyperplanes according to Eq.(\ref{2.5b}), not allowing for
individual time differences between the two particles, it is not explicitly
covariant. In instant-form dynamics, the global propagator $|_0 G_0(K)|_0$
still allows for particle and antiparticle propagation. This is considered
to be a technical disadvantage.

The free resolvent in {\it light-front} coordinates $k_i=(k^-_i:=k^0-k^3\ ,
\ k^+_i:=k^0+k^3 \ , \ \vec k_\perp)$ 
\begin{eqnarray}
\left\langle x_1^{\prime +}x_2^{\prime +}\right| G_0\left|
x_1^+x_2^+\right\rangle &=&-\frac{1}{(2\pi)^2} \int dk_1^-dK^- e^{-\frac{i}{2%
} k_1^-\left( x_1^{\prime +} -x_2^{\prime +}-x_1^++x_2^+\right)} e^{-\frac{i%
}{2} K^-\left(x_2^{\prime +}-x_2^+\right) } \times  \nonumber \\
&&\frac{1}{\widehat k^+_1( K^+-\widehat k^+_1) \left( k_1^--\frac{\widehat{%
\vec k}^2_{1\perp}+m_1^2-io}{\widehat k^+_1}\right) \left(K^--k_1^- -\frac{%
\widehat{\vec k}^2_{2\perp}+m_2^2-io}{K^+-\widehat k^+_1}\right)}
\label{2.8}
\end{eqnarray}
$--$  only its dependence on the individual light-front "times" $%
x^+_i $ is made explicit $--$ reduces, for propagation between the
hyperplanes $x^+$ and $x^{\prime +}$, to 
\begin{eqnarray}
\left\langle x^{\prime +}x^{\prime +}\right| G_0\left| x^+x^+\right\rangle
&=& \int \frac{dK^-}{2\pi} e^{-\frac{i}{2} K^-\left( x^{\prime +}-x^{
+}\right) } \int dk^{\prime -}_1 d k^-_1 \left\langle k_1^{\prime -}\right|
G_0(K)\left|k^-_1\right\rangle \ ,  \label{2.9a} \\
&\equiv& \int \frac{dK^-}{2\pi} e^{-\frac{i}{2} K^-\left( x^{\prime +}-x^{
+}\right) } | G_0(K)| \ .  \label{2.9b}
\end{eqnarray}
In Eq.(\ref{2.9a}) the notation 
\begin{eqnarray}
& & \left\langle k_1^{\prime -}\right| G_0(K)\left|k^-_1\right\rangle \ = 
\nonumber \\
&-&\frac1{2\pi} \frac{\delta \left(k^{\prime -}_1-k^{-}_1\right)}{\widehat k%
^+_1 (K^+-\widehat k^+_1) \left(k_1^--\frac{\widehat{\vec k}^2%
_{1\perp}+m_1^2-io}{\widehat k^+_1}\right) \left(K^--k_1^--\frac{\widehat{%
\vec k}^2_{2\perp}+m_2^2-io}{K^+-\widehat k^+_1} \right)}  \label{2.10}
\end{eqnarray}
is introduced with the abbreviation, 
\begin{eqnarray}
|G_0(K)|:&= & \int dk^{\prime -}_1 dk^{ -}_1 \left\langle k_1^{\prime
-}\right| G_0(K)\left|k^-_1\right\rangle \   \label{2.11a} \\
&=&\frac{i \theta(K^+-\widehat k^+_1) \theta(\widehat k^+_1)} {\widehat{k}%
^+_{1}(K^+-\widehat{k}^+_{1}) \left(K^--\widehat{k}_{1on}^--\widehat{k}%
_{2on}^-+io \right) }  \label{2.11b} \\
&:=& g_0(K) \ ,  \label{2.11c}
\end{eqnarray}
where $K^+\ > \ 0$ can be chosen without any loss of generality.
The matrix element $\left\langle k_1^{\prime -}\right|
G_0(K)\left|k^-_1\right\rangle $ of Eq.(\ref{2.10}), in which only the
dependence on the "dynamic" variable $k^-_1$ is made explicit, still remains
an operator with respect to the "kinematic" variables ($k^+_1 , \vec k%
_{1\perp}$), $\widehat k^-_{1on}=\frac{\widehat{\vec k}_{1\perp}^2+m_1^2}{%
\widehat k^+_1} $ and $\widehat k^-_{2on}= \frac{ (\vec K_\perp-\widehat{%
\vec k}_{1\perp})^2+m_2^2}{K^+-\widehat k^+_1} $. The basis states for the
kinematical light-front variables are defined by 
\begin{eqnarray}
\langle x^-_i \vec x_{i\perp}\left| k^+_i\vec k_{i\perp}\right\rangle=
e^{-\imath(\frac12 k^+_ix^-_i-\vec k_{i\perp} . \vec x_{i\perp})} \ 
\label{basislf}
\end{eqnarray}
and are eigenfunctions of the momentum operators ($\widehat k^+_i , \widehat{%
\vec k}_{i\perp}$) and the free energy operator $\widehat k^-_{ion} $ . The
states $\left|k^+\vec k_\perp \right \rangle$ form an orthonormal and
complete basis, e.g., 
\begin{eqnarray}
\int \frac{dk^+d^2k_\perp}{2(2\pi)^3} \langle x^{\prime-} \vec x%
^{\prime}_\perp\left| k^+\vec k_\perp\right\rangle \langle k^+\vec k%
_\perp\left| x^{-} \vec x_\perp\right\rangle= \delta(x^{\prime -}-x^-)\delta(%
\vec x^{\prime}_\perp-\vec x_\perp) \ .
\end{eqnarray}

In Eq. (\ref{2.11a}) the vertical bar $|$ indicates that the dependence on $%
k^-_1$ is integrated out. The bar on the left of the resolvent represents
integration on $k^-_1$ in the bra-state, the bar on the right in the
ket-state. We shall encounter resolvents in which integration on $k^-_1$ is
done on one side alone, the bar $|$ being placed only on that side. The
operator $g_0(K)$ is three-dimensional and it depends on the kinematic
variables ($k^+_1 , \vec k_{1\perp}$) only. It is a global propagator, since
it mediates between hyperplanes according to Eq.(\ref{2.9b}), not allowing
for individual light-front time differences between the two particles. It
does not possesses explicit covariance 
but is still covariant under light-front
boosts. In light-front dynamics, the global propagator $g_0(K)$ only allows
particle propagation, no antiparticle propagation, due to the
choice of $K^+ > 0$. This is the advantage of light-front dynamics, with
which we work from now on.

The auxiliary four-dimensional resolvent $\widetilde G_0(K)$, introduced in
Eqs. (\ref{2.1})-(\ref{2.3b}) has to be chosen next. We require for $%
\widetilde G_0(K)$: 
\begin{eqnarray}
& &\widetilde G_0(K)|= G_0(K)| \ ,  \label{2.12a} \\
& &|\widetilde G_0(K)= |G_0(K) \ ,  \label{2.12b} \\
& &|\widetilde G_0(K)|= |G_0(K)| \ ,  \label{2.12c}
\end{eqnarray}
and define a three-dimensional transition matrix $t(K)$ through 
\begin{eqnarray}
|\left[\widetilde G_0(K)+\widetilde G_0(K)T(K)\widetilde G_0(K)\right]| =
g_0(K)+g_0(K)t(K)g_0(K) \ .  \label{2.12d}
\end{eqnarray}
In Eqs.(\ref{2.12a})-(\ref{2.12d}) the abbreviation $|$ for integrating out
the $k^-_1$ dependence of operators is used. The conditions (\ref{2.12a})-(%
\ref{2.12d}) are a rather mixed bag. The conditions (\ref{2.12c}) and (\ref
{2.12d}) are physical ones: They require that the global-propagator form of $%
\widetilde G_0(K)$ be the same as for the exact free resolvent $G_0(K)$ and
that the full resolvent of BSE $G_0(K)+G_0(K)T(K)G_0(K)$ can be obtained
from $|\widetilde G_0(K)|$ and the three-dimensional $t(K)$. However, the
two conditions (\ref{2.12c}) and (\ref{2.12d}) do not determine $\widetilde G%
_0(K)$ in full. Our choice is 
\begin{eqnarray}
\widetilde G_0(K):= G_0(K)| g_0^{-1}(K) |G_0(K) \ ,  \label{2.13}
\end{eqnarray}
though $\widetilde G_0(K)=\delta\left(\widehat k^{\prime -}_1 -\frac{K^-}{2}%
\right) g_0(K) \delta\left(\widehat k^-_1-\frac{K^-}{2}\right) $ (and
obvious variants of it) seems to be a legitimate alternative. However, if we
demand that the kernel of the integral equation for the auxiliary transition
matrix, $t(K)$, represents light-front propagation in higher Fock-states,
then the choice is unique. The conditions (\ref{2.12a}) and (\ref{2.12b})
introduce the additional convenience that the auxiliary resolvent be as
close as possible to the exact free one. The auxiliary quantities are
computed in Appendix A.

\section{ Calculational Procedure}

Our calculational procedure amounts to solving three-dimensional integral
equations, whose solutions then yield the covariant results of the BSE by
quadrature.

The four-dimensional transition matrix $T(K)$ is obtained from the
three-dimensional auxiliary one $t(K)$, defined by Eq.(\ref{2.12d}),
through 
\begin{eqnarray}
t(K)=g_0(K)^{-1}|G_0(K)T(K)G_0(K)|g_0(K)^{-1} \ ,  \label{3.1}
\end{eqnarray}
by first iterating the integral equation (\ref{2.1}) once, 
\[
T(K)=W(K)+W(K)\left[\widetilde G_0(K)+ \widetilde G_0(K)T(K)\widetilde G%
_0(K)\right]W(K) \ , 
\]
and then making use of our choice, Eq.(\ref{2.13}), for $\widetilde G_0(K)$
and the result Eq.(\ref{3.1}). The relation between the $T(K)$ and the
auxiliary $t(K)$ is 
\begin{eqnarray}
T(K)=W(K)+W(K)G_0(K)|\left[g_0(K)^{-1}+t(K)\right]|G_0(K)W(K) \ .
\label{3.2}
\end{eqnarray}
The auxiliary transition matrix $t(K)$ itself is obtained by the
three-dimensional integral equation 
\begin{eqnarray}
t(K)=w(K)+w(K)g_0(K)t(K) \ ,  \label{3.3}
\end{eqnarray}
in which the driving term $w(K)$ is derived from the modified
four-dimensional interaction $W(K)$ of Eq.(\ref{2.3a}) according to 
\begin{eqnarray}
w(K):=g_0(K)^{-1}|G_0(K)W(K)G_0(K)|g_0(K)^{-1} \ .  \label{3.4}
\end{eqnarray}
There is an integral equation for $w(K)$ as there is for $W(K)$, but we do
not give it here. 
We hope that, through our choice (\ref{2.13}) for $\widetilde G_0(K)$,
a few terms of the expansion Eq.(\ref{2.3a}),
 of $W(K)$ in powers of $V(K)$ will
dynamically suffice to yield the full result of BSE with satisfactory
accuracy. The numerical example of Sect. IV where  rapid convergence 
of $w(K)$ is seen, demonstrates the validity of this expectation.

If the transition matrix $T(K)$ of the BSE has a bound-state pole at total
four momentum $K_B$, $K^2_B=M^2_B$, the auxiliary three-dimensional
transition matrix $t(K)$ also has a bound-state pole at exactly the same $%
K_B $, according to Eq.(\ref{3.1}), with the residue $\left|
\gamma_B\right\rangle $ being the solution of the homogeneous
three-dimensional equation 
\begin{eqnarray}
\left| \gamma_B\right\rangle= w(K_B)g_0(K_B) \left| \gamma _B\right\rangle \
,  \label{3.5}
\end{eqnarray}
corresponding to the inhomogeneous one, Eq.(\ref{3.3}). From $\left| \gamma
_B\right\rangle$, the residue $\left| \Gamma _B\right\rangle$ of BSE can be
recovered according to Eq.(\ref{3.2}) 
\begin{eqnarray}
\left| \Gamma_B\right\rangle= W(K_B)G_0(K_B)| \left| \gamma _B\right\rangle
\end{eqnarray}
as well as the bound state $\left| \Psi_B\right\rangle$ of BSE, i.e. 
\begin{eqnarray}
\left| \Psi _B\right\rangle &=& G_0(K_B) W(K_B) G_0(K_B)|\left| \gamma
_B\right\rangle \ ,  \label{3.6a} \\
\left| \Psi _B\right\rangle &=&\left[
1+\left(G_0(K_B)-G_0(K_B)|g_0(K_B)^{-1}|G_0(K_B)\right) W(K_B)\right]
G_0(K_B)|\left| \gamma _B\right\rangle \ .  \label{3.6b}
\end{eqnarray}
For the form Eq.(\ref{3.6b}) of the bound state, 
the condition Eq.(\ref{3.5}) $%
\left| \gamma_B\right\rangle- w(K_B)g_0(K_B) \left| \gamma _B\right\rangle
=0 $ is used. The step from the three-dimensional residue $\left| \gamma
_B\right\rangle $ to the four-dimensional bound state $\left|\Psi_B\right%
\rangle $ appears predominantly a kinematic one, effected by the operator $%
G_0(K_B)|$ . Only the second term in Eq.(\ref{3.6b}) depends on the
interaction, and it is expected to be a small correction.

The four-dimensional bound state $\left|\Psi_B\right\rangle $ is related to
the auxiliary three-dimensional $\left|\phi_B\right\rangle $, defined by 
\begin{equation}
\left|\phi_B\right\rangle:=g_0(K_B)\left|\gamma_B\right\rangle
\end{equation}
and satisfying 
\begin{equation}
\left|\phi_B\right\rangle=g_0(K_B)w(K_B)\left|\phi_B\right\rangle
\ ,
\end{equation}
in an obvious way by 
\begin{eqnarray}
\int dk^-_1 \left\langle k^-_1 | \Psi _B\right\rangle =\left| \phi_B
\right\rangle \ .  \label{3.6c}
\end{eqnarray}
The result Eq.(\ref{3.6c}) follows immediately from 
Eq.(\ref{3.6b}). The auxiliary
bound-state wave-function $\left| \phi_B \right\rangle$ is the projection of
the bound-state $\left|\Psi_B \right\rangle$ of BSE to equal light-front
individual times $x^+_i=x^+$, taken on the hyperplane $x^+=0$.

The bound-state $\left| \Psi_B \right\rangle$ of BSE and its
three-dimensional auxiliary version $\left| \phi_B \right\rangle$ still have
to be normalized. If the dependence on $K$ of the original interaction $V(K)$ 
is weak, i.e., $(V(K)-V(K_B))/(K^2-K^2_B)\simeq 0$ and if furthermore the
interaction-dependent term in the step from $\left| \phi_B\right\rangle$ to $%
\left|\Psi_B\right\rangle$ according to Eq.(\ref{3.6b}) is small, i.e., $%
\left| \Psi _B\right\rangle \simeq G_0(K_B)|g_0(K_B)^{-1}\left| \phi
_B\right\rangle $, then 
\begin{eqnarray}
& &\lim_{K^2\rightarrow K^2_B} \left\langle \Psi_B \left| \frac{%
G_0(K)^{-1}-G_0(K_B)^{-1}}{K^2-K^2_B} -\frac{V(K)-V(K_B)}{K^2-K^2_B} \right|
\Psi _B\right\rangle \simeq  \nonumber \\
& &\lim_{K^2\rightarrow K^2_B} \left\langle \Psi_B \left| \frac{%
G_0(K)^{-1}-G_0(K_B)^{-1}}{K^2-K^2_B}\right| \Psi _B\right\rangle \simeq 
\nonumber \\
& &\lim_{K^2\rightarrow K^2_B} \left\langle \phi_B \left| \frac{%
g_0(K)^{-1}-g_0(K_B)^{-1}}{K^2-K^2_B}\right| \phi _B\right\rangle = 1 \ .
\label{3.6f}
\end{eqnarray}
For any further applications, i.e., for predicting physical observables, we
now have two equally valid options. We may either work with covariant
operators using the bound state $\left|\Psi_B\right\rangle$ and/or the
transition matrix $T(K)$ of the BSE or we may derive effective operators
suited for the context of the auxiliary three-dimensional bound state $%
\left|\phi_B\right\rangle$ and/or the auxiliary three-dimensional transition
matrix $t(K)$. We give an example of each of the possible strategies:

We use the eletroweak current ${\cal J}^\mu(Q)$ as example and assume that
it connects an initial bound state $\left| \Psi_{Bi}\right\rangle$ to a
final one $\left| \Psi_{Bf}\right\rangle$ in an elastic process.
We take ${\cal J}^\mu(Q)$ to be the current 
appropriate for the hadronic field theory with
four-momentum transfer $Q=K_{Bf}-K_{Bi}$. The matrix element for describing
the process $\left\langle \Psi_{Bf}\right|{\cal J}^\mu(Q)\left|
\Psi_{Bi}\right\rangle$ can be obtained from the three-dimensional bound
state $\left|\phi_{B}\right\rangle$ by 
\begin{eqnarray}
\left\langle \Psi_{Bf}\right|{\cal J}^\mu(K_{Bf}-K_{Bi}) \left|
\Psi_{Bi}\right\rangle = \left\langle\phi_{Bf}\right|j^\mu(K_{Bf},K_{Bi})
\left| \phi_{Bi}\right\rangle \ , \label{3.7a}
\end{eqnarray}
with the effective current in three-dimensional space 
\begin{eqnarray}
& &j^\mu(K_f,K_i):=g_0(K_f)^{-1}|G_0(K_f)\left[1 +W(K_f)\left(G_0(K_f)-
G_0(K_f)|g_0(K_f)^{-1}|G_0(K_f)\right)\right]  \nonumber \\
& &{\cal J}^\mu(K_f-K_i) \left[1 +\left(G_0(K_i)-
G_0(K_i)|g_0(K_i)^{-1}|G_0(K_i)\right)W(K_i)\right] G_0(K_i)|g_0(K_i)^{-1} \ .
  \label{3.7b}
\end{eqnarray}
For the relation between the bound states $\left| \Psi_{B}\right\rangle$ and 
$\left| \phi_{B}\right\rangle$, Eq.(\ref{3.6b}) is used, which separates the
kinematic and dynamic, i.e., interaction-dependent, steps in that relation
from each other. The bound state has to be calculated for the initial and
final four-momenta $K_{Bi}$ and $K_{Bf}$. The effective current $%
j^\mu(K_f,K_i)$ is predominantly derived
kinematically from the covariant one
through $g_0(K_f)^{-1}|G_0(K_f){\cal J}^\mu(K_f-K_i)G_0(K_i)|g_0(K_i)^{-1}$
but it also depends on the interaction $W(K)$ of Eq.(\ref{2.3a}). If $W(K)$
is not computed in full, but only expanded up to a certain order in the
original interaction $V(K)$ of the BSE, the effective current should be
expanded consistently up that order.

\section{A Numerical Test Case}

We use the bound state of a schematic two-boson system as a test case of
the power of the suggested numerical technique. The employed interaction
Lagrangian is 
\begin{eqnarray}
{\cal L}_I=g_S\phi _1^{\dagger }\phi _1\sigma +g_S\phi _2^{\dagger }\phi
_2\sigma ,  \label{li}
\end{eqnarray}
where the bosons with fields $\phi_1$ and $\phi_2$ have masses $m_1$ and $%
m_2 $, which we take to be equal,
$m_1=m_2=m$, and the exchanged boson with field $\sigma$ has
mass $\mu$. The coupling constant is $g_S$.

Using standard techniques in Euclidean space, the homogeneous BSE is solved
for the bound-state vertex $\left| \Gamma_{B}\right\rangle$ in the ladder
approximation, i.e., 
\begin{equation}
\langle k^{\prime}_{1}\left| \Gamma_{B}\right\rangle =ig_S^2\int \frac{d^4k_1%
}{(2\pi )^4} \frac{\langle k_{1}\left| \Gamma_{B}\right\rangle} {\left(
(k^\prime_1-k_1)^2-\mu ^2+i\varepsilon \right) \left( k_1^2-m^2+i\varepsilon
\right) \left( (K_B-k_1)^2-m^2+i\varepsilon \right) } \ .  \label{bs}
\end{equation}

The solution is calculated in the two-particle c.m. system, i.e., for $%
K_B=(M_B,\vec 0)$, and for the ratio of masses $\mu/m=0.5$. Requiring the
bound state mass to have a particular value $M_B$ fixes the coupling
constant $g_S$. 
The four-dimensional bound-state vertex $\langle k_{1}\left|
\Gamma_{B}\right\rangle$ depends on all Euclidean four components of the
momentum $k_1$ of boson 1. The exact four-dimensional bound state is
obtained according to Eq.(\ref{2.2b}).
However, the representation of the vertex
and bound state in terms of Minkowski momenta is difficult. We do not
attempt it.

In contrast, the four-dimensional bound-state obtained by the numerical
technique suggested in Sect.III is available in Minkowski space. We
calculate it only approximately by using for the driving term $w(K_B)$ of
the auxiliary three-dimensional equation Eq.(\ref{3.5}),
 an expansion in orders
of the interaction $V(K)$ of BSE in Eqs. (\ref{2.3b}) and (\ref{3.4}), i.e.,
in powers of the coupling constant $g_S$ of the interaction Lagrangian (\ref
{li}). We use the approximation up to the second and fourth powers of $g_S$,
i.e., $w(K_B)\simeq w^{(2)}(K_B)$ and $w(K_B)\simeq
w^{(2)}(K_B)+w^{(4)}(K_B) $. In a time-ordered view, the BSE allows for an
exchange of an infinite number of $\sigma$ bosons in stretched
configurations. In contrast, the approximative $w^{(2)}(K_B)$ allows only
for one exchange (Fig.1a), while $w^{(4)}(K_B)$ allows for two (Fig.1b). 
The analytic forms of $w^{(2)}(K_B)$ and 
$w^{(4)}(K_B)$ are given in Appendices B and C. 
The explicit forms of the homogeneous 
integral equation for $\left| \gamma_B\right\rangle$, Eq.(\ref{3.5}),
for the above approximations in the driving term are given in 
Appendix D. In order to make a
comparision with the exact bound state we study the 
projected forms of the bound states, i.e., 
\begin{eqnarray}
f_{exact}(\sqrt{\vec k^2_{1\perp}}):&=& \int dk^-_1 dk^+_1 \langle
k_{1}\left| \Psi_{B}\right\rangle  \label{fperp} \\
&=&2\int dk^0_1 dk^3_1 \langle k^0_{1} \vec k_{1\perp} k^3_1 \left|
G_0(K_B)\right|\Gamma_{B}\rangle \ .  \nonumber
\end{eqnarray}
\begin{eqnarray}
f^{(n)}_{app}(\sqrt{\vec k^2_{1\perp}})&=& \int dk^+_1 \langle k^+_{1}\vec k%
_{1\perp}\left| \phi_{B}\right\rangle^{(n)}_{app}  \label{fapp} \\
&=&\int dk^+_1 \langle k^+_{1}\vec k_{1\perp} \left|
g_0(K_B)\right|\gamma_{B}\rangle_{app}^{(n)} \ .  \nonumber
\end{eqnarray}
The superscripts $(n)$ in Eq.(\ref{fapp}) indicate the power of the coupling
constant up to which the approximation is carried, i.e., $w(K_B)\simeq
\sum_{i=2}^n w^{(i)}(K_B)$. The comparision between exact and approximate
results is carried out on two levels:

In Fig. 2  the relation between $g_S$ and $M_B$ is tested for $\mu=0.5m$
against the four-dimensional results.
Whereas the
exact relation is already satisfatory reproduced by the approximation based
on $w^{(2)}(K_B)$, the approximation based on $w^{(2)}(K_B)+w^{(4)}(K_B)$ 
improves the agreement.

In Figs. 3 and 4, the projected bound-states $f(\sqrt{\vec k^2_{1\perp}})$ are
compared for two cases. 
In the first case $M_B=0$, i.e., the binding is very strong. It
is of the order of the masses of the interacting particles as encountered in
quark systems. In the other case $M_B=1.98 m$, i.e., the binding is very
weak. It is only  2\% of the masses of the interacting
particles, as encountered in nuclear systems. In both cases the approximation
based on $w^{(2)}(K_B)$ is already quite accurate. The improvement due to
the inclusion of $w^{(4)}(K_B)$ is particularly noticeable for the case of
strong binding.

The fact that a low-order approximation of $w^{(n)}(K_B)$ works surprisingly
well is a virtue of light-front dynamics. It is well known that the 
analogous approximation scheme in instant-form dynamics has much poorer
convergence properties with respect to the number of exchanged $\sigma$
bosons \cite{nico}.

\section{Extension to Fermions}

The free resolvent which propagates two fermions disconnectedly contains
self-energy corrections as in the case of bosons. They are usually left out
of the ladder approximation of  interaction. The two-fermion
free resolvent then takes the form which we immediately rewrite conveniently 
as
\begin{eqnarray}
G_0^F&=&\frac{\widehat {\rlap\slash k}_1 +m_1}{\widehat k_1^2-m_1^2} \frac{%
\widehat{\rlap\slash k}_2 +m_2}{\widehat k_2^2-m_2^2}  \label{e.1a} \\
G_0^F&=&\Delta G_0^F + \left(\widehat {\rlap\slash k}_{1on} +m_1\right)
\left(\widehat {\rlap\slash k}_{2on} +m_2\right)G_0 \ ,  \label{e.1b}
\end{eqnarray}
where $\widehat k^-_{1on}=\frac{\widehat{\vec k}_{1\perp}+m_1^2} {\widehat k%
^+_1}$ and $\widehat k^-_{2on}=\frac{\widehat{\vec k}_{2\perp}+m_2^2}{%
\widehat k^+_2}$. In Eq.(\ref{e.1b}) $G_0$ is the covariant propagator the
paper has worked with in the conceptual development untill now. Furthermore,
Eq.(\ref{e.1b}) is the definition of $\Delta G_0^F$ which contains -- except
for the particular spin-dependent operators $\left(\widehat {\rlap\slash k}%
_{1on} +m_1\right)$ and $\left(\widehat {\rlap\slash k}_{2on} +m_2\right)$
that commute with $G_0$ -- all particular divergences and subtleties connected
with the fermion motion. The operator 
\begin{eqnarray}
\Delta_0^F= \frac{\gamma^+_1}{2k_1^+} \frac{\widehat{\rlap\slash k}_{2on}
+m_2}{\widehat k_2^2-m_2^2}+ \frac{\widehat{\rlap\slash k}_{1on} +m_1}{%
\widehat k_1^2-m_1^2} \frac{\gamma^+_2}{2k_2^+} +\frac{\gamma^+_1}{2k_1^+} 
\frac{\gamma^+_2}{2k_2^+}  \label{inst}
\end{eqnarray}
carries the instantaneous part of the fermion propagators in light-front
time. Its is singular under $k^-_1$ integration. We therefore suggest the
following strategy for fermions: We apply the reduction
to an auxiliary resolvent 
$\widetilde G_0$ twice, using the apparatus of Sects. I and II. The operator
dependence on the total two-fermion four momentum $K$ is factored out as
there. All operators become then parametrically dependent on $K$.

In the first step, the two-fermion resolvent $\left(\widehat {\rlap\slash k}%
_{1on} +m_1\right) \left(\widehat {\rlap\slash k}_{2on} +m_2\right) G_0(K)$
is introduced instead of $G_0^F(K)$. We use formulae (\ref{2.1})-(\ref{2.3b}%
) to do this. All the physics of anomalous two-fermion propagation is
contained in the new effective interaction $W(K)$ of Eq.(\ref{2.3a}). Thus,
one arrives at a new BSE, corresponding to Eq.(\ref{1.1}) after reduction
with respect to $K$, with the four dimensional resolvent $\left(\widehat {%
\rlap\slash k}_{1on} +m_1\right)\left(\widehat {\rlap\slash k}_{2on}
+m_2\right)G_0(K)$ and the new interaction. The resulting two-fermion
equation is now solved with the technique as developed for two bosons. This
is possible due to the fact that the spin-dependent operator $\left(%
\widehat {\rlap\slash k}_{1on} +m_1\right)\left(\widehat {\rlap\slash k}%
_{2on} +m_2\right)$ also commutes with the auxiliary one $\widetilde G%
_0(K)=G_0(K)|g_0(K)^{-1}|G_0(K)$, i.e., 
\begin{eqnarray}
\left(\widehat {\rlap\slash k}_{1on} +m_1\right)\left(\widehat {\rlap\slash k%
}_{2on} +m_2\right) \widetilde G_0(K)=\widetilde G_0(K) \left(\widehat {%
\rlap\slash k}_{1on} +m_1\right) \left(\widehat {\rlap\slash k}_{2on}
+m_2\right) \ .  \label{e.2}
\end{eqnarray}

This idea will not be further developed in this paper, but indicates
that the scope of the method extends beyond the two-boson system.

\section{ Relation to Light-Front Quantum Mechanics}

Sects. I-IV used the notion of a bound state, but scattering states were not
introduced. The later could have been introduced in 
the BSE (\ref{1.1}) as well as in
the auxiliary three-dimensional equation (\ref{3.3}) for $t(K)$ with the
global propagator $g_0(K)$. Given an initial two-particle plane-wave state $%
\left| k^+_1\vec k_{1\perp} K_{on}\right\rangle $ with total momentum $%
K_{on} $ and light-front "energy" $K^-_{on}=\frac{\vec k_{1\perp}^2+m_1^2}{%
k^+_1} +\frac{(\vec K_\perp-\vec k_{1\perp})^2+m_2^2}{K^+-k^+_1}$, one may
define the corresponding three-dimensional scattering state $\left|
\phi^{(+)}( k^+_1\vec k_{1\perp} K_{on})\right\rangle $ with outgoing
light-front boundary conditions as the solution of a standard
Lippman-Schwinger type of equation, i.e., 
\begin{eqnarray}
\left| \phi^{(+)}( k^+_1\vec k_{1\perp} K_{on})\right\rangle =\left| k^+_1%
\vec k_{1\perp} K_{on}\right\rangle +g_0(K_{on})w(K_{on}) \left| \phi^{(+)}(
k^+_1\vec k_{1\perp} K_{on})\right\rangle  \label{5.1}
\end{eqnarray}
with four-momentum $K_{on}=(K^-_{on},K^+,\vec K_\perp)$. The relation to the
auxiliary transition operator $t(K)$ is obvious, 
\begin{eqnarray}
t(K_{on})\left| k^+_1\vec k_{1\perp} K_{on}\right\rangle = w(K_{on}) \left|
\phi^{(+)}( k^+_1\vec k_{1\perp} K_{on})\right\rangle \ .  \label{5.2}
\end{eqnarray}
Furthermore, it satisfies the homogeneous equation 
\begin{eqnarray}
\left[g_0(K_{on})^{-1}-w(K_{on})\right] \left|\phi^{(+)}( k^+_1\vec k%
_{1\perp} K_{on})\right\rangle =0  \label{5.3a}
\end{eqnarray}
in the same way as the auxiliary bound state $\left| \phi_B\right\rangle $
of Eq.(\ref{3.5}) does, i.e., 
\begin{eqnarray}
\left[g_0(K_B)^{-1}-w(K_B)\right] \left| \phi_B\right\rangle =0  \label{5.3b}
\end{eqnarray}
Eqs. (\ref{5.3a}) and (\ref{5.3b}) formally look like the eigenvalue
equations of quantum mechanics with the only difference being that the
two-particle interaction $w(K)$ depends on the eigenvalue. Untill now the
relationship to quantum mechanics has indeed been entirely formal. 
The states $\left| \phi_B\right\rangle$ and $\left|\phi^{(+)}( k^+_1\vec k%
_{1\perp} K_{on})\right\rangle $ and the corresponding transition matrix
have significance only as quantities from which the solutions of the BSE
can be obtained with comparative ease. On the other hand, at this stage a
quantum-mechanical description of the two-particle system can be given which
corresponds dynamically to the underlying field-theoretic one, though it is by
no means equivalent to it. Quantum-mechanical two-particle states $\left|
\varphi\right\rangle $ are required to satisfy the eigenvalue equation for
the squared mass operator,
\begin{eqnarray}
\left[M^2_0+v(K^+,\vec K_\perp)\right] \left| \varphi\right\rangle
=M^2_B\left| \varphi\right\rangle \ ,  \label{5.4}
\end{eqnarray}
where the squared free-mass operator is 
\begin{equation}
M^2_0=\frac{\widehat{\vec k}^2_{1\perp}+m_1^2}{\hat x}+\frac{(\vec K_\perp -%
\widehat{\vec k}_{1\perp})^2+m_2^2}{1-\hat x} \ ,  \label{m0}
\end{equation}
and $\hat x=\frac{\hat k^+}{\hat K^+}$. The states are elements of a
Hilbert-space spanned by the free-particle on-mass-shell basis states.
Boundary conditions must be imposed on the solutions of Eq.(\ref{5.4}) in
order to make them acceptable. Bound-state and scattering state
solutions to the mass squared operator equation exist and are orthonormalized.
The orthonormalization for scattering states is of the $\delta$-function type.
The states have a probability interpretation. The quantum mechanical
bound-state normalization is 
\begin{equation}
\langle\varphi_{B }|\varphi_{B}\rangle=1 \ .
\end{equation}

The two-particle potential $v(K^+,\vec K_\perp)$ is independent of the
eigenvalue $K^-$, the eigenvalue $K^-_B$ to be calculated for the bound
state and the eigenvalue $K^-_{on}$ prescribed for the scattering states;
the potential is hermitian it is instantaneus in light-front time; it
conserves the kinematic components $(K^+,\vec K_\perp)$ of the total
two-particle four-momentum $K$. In quantum mechanics $v(K^+,\vec K_\perp)$
may be parametrized by fitting it to observables. If contact is attempted to
a corresponding field theory a standard form of identification is 
\begin{eqnarray}
\langle k^{\prime +}_1 \vec k^{\prime}_{1\perp} \left|v(K^+,\vec K_\perp)
\right|k^{ +}_1 \vec k_{1\perp}\rangle:= i \sqrt{\frac{K^+}{k^{\prime
+}_1(K^+-k^{\prime +}_1)}} \langle k^{\prime +}_1 \vec k^{\prime}_{1\perp}
\left|w(K_v) \right|k^{ +} \vec k_\perp\rangle \sqrt{\frac{K^+}{%
k^+_1(K^+-k^+_1)}}  \label{5.5a}
\end{eqnarray}
with 
\begin{eqnarray}
K_v=(\frac12 K^{\prime -}_{on}+\frac12 K^{ -}_{on}, K^+,\vec K_\perp) \ .
\label{5.5b}
\end{eqnarray}
The defined relativistic quantum mechanical interaction is 
cooked in the framework of light-front dynamics.
The value $K^-_{on}$ is defined in the context of Eq.(\ref{5.4}). This
choice guarantees that the S-matrix calculated field-theoretically to first
order in $w(K)$ and calculated quantum-mechanically to first order in $v(K^+,%
\vec K_\perp)$ are identical. The S-matrix carries a $\delta$-function for
light-front energy $K^-$ between initial and final states. The definition 
of Eq.
(\ref{5.5a}) removes that $\delta$-function from $v(K^+,\vec K_\perp)$ and
allows for general off  $K^-$-shell matrix elements. Thus, Eq.(\ref{5.5b}%
) implies a very particular off-shell extension. This procedure of
identification -- it is no derivation -- is standard for the instant-form of
quantum mechanics, e.g., when the one-boson exchange potential between
nucleons is introduced. This paper extends that procedure to light-front
quantum mechanics. Furthermore, the potential is usually defined in the
two-particle c.m. system, i.e., for $\vec K_\perp=0$, and is considered
unchanged in moving systems, i.e., independent of $\vec K_\perp$ and $K^+$.

The identification (\ref{5.5a}) motivates a quantum-mechanical potential. It
does not attempt to derive it. The goal of 
the identification is to simulate exact
solutions of the BSE in best accord with a chosen physics criterion, quantum
mechanics description has different objectives than matching a field-theory
result. It rather attempts to describe many-particle systems
with the same rules once it has done
so satisfactorily for the two-particle system with the same rules. Thus,
when the quantum-mechanical potential cannot be derived completely, as is the
case in hadronic physics, the potential is tuned to known experimental
properties of the two-particle system and then considered a vehicle which
carries that two-particle information to many-particle systems. Despite the
particular many-particle aspect of quantum mechanics, a study of its
predictive quality even for the two-particle system is interesting. Figs. 2-4
perform such study for the two-boson system of Sect.IV. 
The bound state constitutes an
especially stringent test. For the instantaneous choice, 
the approximation, $K^-=K^-_{on}
\ > \ m_1+m_2$ in the interaction in the c.m. system is quite  severe,
because
in this case  field theory requires $K^-=K^-_B \ < \ m_1+m_2$. The
relation between the coupling constant $g_S$ and 
the bound-state mass $M_B$ and the
dependence of the bound-state wave-function $f(q)$ on
the momentum $q=\sqrt{\vec k_\perp^2}$ are compared in the field-theoretic and
quantum-mechanical descriptions. Results are studied for the approximations $%
w(K)\simeq w^{(2)}(K)$ and $w(K)\simeq w^{(2)}(K)+w^{(4)}(K)$ up to second
order and fourth-order in the coupling constant $g_S$. The
quantum-mechanical binding energy and wave-function preserve most
field-theoretic characteristics, expectedly better in the case of small
binding rather tha in the case of strong binding. The quantum-mechanical
choice of the potential is usually based on the one-boson exchange, i.e., on
the approximation $w(K)\simeq w^{(2)}(K)$. We are happy to find that this
identification accounts better for the field-theoretic results than the
choice based on $w(K)\simeq w^{(2)}(K)+w^{(4)}(K)$.

Instead of solving Eq.(\ref{5.4}), its formal identity with the energy
eigenvalue problem for a nonrelativistic hamiltonian is often exploited 
\cite{frank} and $\left|\varphi_{B} \right\rangle $ is applied directly
in the framework of light-front quantum-mechanics.

The response of the quantum-mechanical system to an electromagnetic probe is
given by a four-vector current $j^\mu_v(K^{\prime +}-K^+,\vec K%
^{\prime}_\perp-\vec K_\perp)$ which,  as the
quantum-mechanical potential is a three-dimensional operator and
 it depends on the three-dimensional momentum
transfer $(Q^+,\vec Q_\perp)=(K^{\prime +}-K^+,\vec K^{\prime}_\perp-\vec K%
_\perp)$. As in the case of the potential, contact can be attempted with the
corresponding field theory. A possible identification is 
\begin{eqnarray}
& &\langle k^{\prime +}_1 \vec k^{\prime}_{1\perp} \left|j^\mu_v(K^{\prime
+}-K^+,\vec K^{\prime}_\perp-\vec K_\perp) \right|k^{ +}_1 \vec k%
_{1\perp}\rangle :=  \nonumber \\
& & \sqrt{\frac{K^+}{k^{\prime +}_1(K^+-k^{\prime +}_1)}} \langle k^{\prime
+}_1 \vec k^{\prime}_{1\perp} \left|j^\mu(K^\prime _v-K_v) \right|k^{ +}_1 
\vec k_{1\perp}\rangle \sqrt{\frac{K^+}{k^+_1(K^+-k^+_1)}}  \label{curqm}
\end{eqnarray}
with 
\begin{eqnarray}
K_v&=&(K_{on},K^+,\vec K_\perp)  \nonumber \\
K^\prime_v&=&(K^\prime_{on},K^{\prime +},\vec K^\prime_\perp) \ .
\end{eqnarray}
The field-theoretic $j^\mu(K^\prime-K)$ is the one of Eq.(\ref{3.7b}) in
Sect. III. It contains the field-theoretic interaction in the form of $w(K)$.
The quantum-mechanical current $j^\mu_v(K^{\prime +}-K^+,\vec K%
^{\prime}_\perp-\vec K_\perp)$, is derived in 
the special case of elastic scattering
between bound states. Thus, the identification 
of Eq.(\ref{curqm}) is not consistent
with the choice of Eq.(\ref{5.5a}), which guaranteed the agreement of the
field-theoretic and the quantum-mechanical S-matrix in first order in the
interaction. Nevertheless, the quantum-mechanical current $j^\mu_v(K^{\prime
+}-K^+,\vec K^{\prime}_\perp-\vec K_\perp)$ can be meaningfully studied and
separated into interaction-free single particle and interaction-dependent
two-particle pieces. Thus, the definition of
Eq.(\ref{curqm})  implicitly contains  a
possible quantum-mechanical definition of an interaction-dependent
two-particle current. At this stage, the standard definition based on the
identification of the S-matrix could also be given \cite{ralf}. 
It also exploits
the formal identity of the eigenvalue problem with a nonrelativistic
hamiltonian with equations like (\ref{5.4}), but it then identifies the
nonrelativistic bound state with the solution $\left|\phi_B \right\rangle$
of Eq.(\ref{5.3b}), the auxiliary field-theoretic bound state for the BSE.
Thus, the calculation of the electromagnetic deuteron form factors in Ref.%
\cite{ralf} is performed in the field-theoretic spirit of Eq.(\ref{5.3b}). The
two-particle current operators of pion range in Ref.\cite{ralf} should not
be confused with the quantum-mechanical interaction-dependent two-particle
currents of this section.

\section{Conclusion}

The paper suggests a calculational procedure for solving the BSE 
with comparative
ease and in principle, with any desired accuracy. The procedure is based on
an auxiliary three-dimensional integral equation, in the framework of
light-front dynamics, whose solution then yields the result of the BSE by
quadrature. The intermediate auxiliary quantities do not display covariance;
covariance is restored in the final step to the full result of BSE.

The calculational procedure is exact, but it also offers an efficient
approximative scheme: Only particles propagate. Antiparticles do not.
Antiparticle propagation is relegated to the effective interaction. The
convergence with repect to the number of exchanged particles mediating
the interaction appears to be rapid. Though only an 
indication of that fact comes
from the simple test case of a BSE bound state in ladder approximation, it
gets supported by the similar result of Ref.\cite{nico} for the
corresponding scattering amplitude. Calculational improvements are possible
in a systematic manner. Thus, as a further and physically more interesting
consequence, the solution of the BSE for bound state and scattering up to
fourth order in the coupling constant, i.e., in ladder and crossed ladder
approximation and with the inclusion of self-energy corrections is obtained
based on a simplifying three-dimensional calculational procedure. The
procedure capitalizes on beneficial properties of light-front dynamics. It
should be an interesting alternative to the Gross approach \cite{gross}
which is also three-dimensional and which has been suggested to include the
cross-ladder exchanges approximately.

The calculational procedure is general, though it is given in this paper for
an interacting two-boson system only. The ideas needed for an extension to
fermions are developed but important technical details have not yet been worked
out and unforeseen difficulties may still arise. The problem of
rotational invariance in light-front dynamics will become especially acute
for fermions when spin and orbital angular momentum are to be coupled. The
auxiliary three-dimensional quantities will then be hampered by their lack of
rotational invariance. We strongly believe however,
that the final step to the
covariant result of BSE will overcome that difficulty.

The auxiliary thee-dimensional quantities, i.e., the operators and
equations, that mediate the solution of the BSE, are close in spirit to
relativistic quantum mechanics. The paper also discusses this relation.
First, only particles and not antiparticles, propagate in the three-dimensional
equations and in quantum mechanics. Second, the quantum-mechanical
interaction is an instantaneous potential, the corresponding interaction $%
w(K)$ in the three-dimensional equation is not. However, this paper finds
that the instantaneous choice for the potential does not distort the physics
of the underlying field theory. Thus, the relation between quantum mechanics
and field theory can be made close. However, compared to field theory,
quantum mechanics has the virtue of an instant extension to many-particle
systems: Barring small corrections due to many-particle forces and the
quantum mechanical interaction is additive in the instantaneous pairwise
potentials. In fact, the conceptual strategy of quantum mechanics often is
to tune away
shortcomings of the chosen instantaneous potential by
adjusting undetermined phenomenological parameters to vital known
experimental properties of the considered two-particle system. In this way
the potential carries the accepted knowledge on the two-particle system over
to many-particle systems.

The paper left open the relationship of the theoretical apparatus 
developed to realistic physics problems. 
We have in mind applications to hadronic
and subhadronic systems. The concept of light-front wave-functions was
applied in the context of nuclear physics to describe the deuteron\cite
{frank} and the discussion of its properties in the light-front continues
to the present \cite{karmanov}. The BSE is supposed to yield 
bound states and the
scattering amplitude for those two-particle systems. In contrast, the
response of such a two-particle system towards an eletroweak probe is
considered in perturbation theory. The required matrix element is determined
by the field-theoretic current between states of the BSE. The paper offers
two equivalent routes for calculation: Either the covariant states of BSE
are constructed and then used in their four-dimensional forms or the
field-theoretic current is reduced to a three-dimensional one, consistent
with the auxiliary three-dimensional ones. Again, the latter calculational
scheme is close to the quantum-mechanical one in spirit. The definition of
two-particle exchange currents for the use in quantum mechanics is sketched.

\section*{Acknowledgments}

The authors acknowledge the partial support from the CAPES/DAAD/Probral
project 015/95. Part of this work was done during the visits of, T.F.
and P.U.S.. They thank the hospitality at the respective host institutions.
Furthermore, the work was supported by a Brazilian graduate-student grant
(J.H.O.S.) from FAPESP and by  research grants (B.V.C. and T.F.) from CNPq
and FAPESP.

\newpage
\appendix

\section{Evaluation of auxiliary quantities}

\setcounter{equation}{0} 

The operators $G_0(K)|g_0(K)^{-1}$ and $g_0^{-1}|G_0(K)$ connect
three-dimensional and four-dimensional basis states. The two operators are
related by conjugation; we therefore discuss only one, i.e., $%
G_0(K)|g_0(K)^{-1}$.

The momentum space matrix elements of $G_0(K)|g_0(K)^{-1}$ for $K^{+}>0$,
are 
\begin{eqnarray}
&&\langle k_1^{\prime -}k_1^{\prime +}\vec{k}_{1\perp }^{\prime }\left|
G_0(K)|g_0(K)^{-1}\right| k_1^{+}\vec{k}_{1\perp }\rangle =\frac i{2\pi }%
\frac{\delta \left( k_1^{\prime +}-k_1^{+}\right) \delta \left( \vec{k}%
_{1\perp }^{\prime }-\vec{k}_{1\perp }\right) }{\left( k_1^{\prime -}-\frac{%
\vec{k}_{1\perp }^{\prime 2}+m_1^2-io}{k_1^{\prime +}}\right) }\times 
\nonumber \\
&&\frac{\left( K^{-}-k_{1on}^{-}-k_{2on}^{-}+io\right) \theta
(K^{+}-k_1^{+})\theta (k_1^{+})}{\left( K^{-}-k_1^{\prime -}-\frac{(\vec{K}%
_{\perp }-\vec{k}_{1\perp }^{\prime })^2+m_2^2-io}{K^{+}-k_1^{^{\prime }+}}%
\right) } \ . \label{a1}
\end{eqnarray}
When the avaliable light-front ``energy'' $K^{-}$ is not on-shell, i.e., $%
K^{-}\neq k_{1on}^{-}+k_{2on}^{-}$, the evaluation of the matrix element in
Eq.(\ref{a1}) is standard. The two singular propagators $\left( k_1^{\prime
-}-\frac{\vec{k}_{1\perp }^{^{\prime }2}+m_1^2-io}{k_1^{\prime +}}\right)
^{-1}$ and $\left( K^{-}-k_1^{\prime -}-\frac{(\vec{K}_{\perp }-\vec{k}%
_{1\perp }^{\prime +})^2+m_2^2-io}{K^{+}-k_1^{\prime +}}\right) ^{-1}$ can
be rewritten as a $\delta $-function and principal-part singularity; 
integration on $k_1^{\prime -}$ can be carried out with usual
techniques.

A problem arises, when the avaliable light-front ``energy'' $K^{-}$ is
on-shell, i.e., $K^{-}=K_{on}^{-}=k_{1on}^{-}+k_{2on}^{-}$. Without losing
generality, we will have suppose that $K^{+}>0$ and $k_1^{+}>0$. Then, $%
K^{-}-k_{1on}^{-}-k_{2on}^{-}+io=+io$ and the limiting process of going to
the real axis must be performed with care. 
However, in this situation the matrix
element will always be integrated with
respect to $k_1^{\prime -}$,  over a function $f(k_1^{\prime -})$ 
still to be determined and, unfortunately with
unknown analyticity properties, i.e., 
\begin{eqnarray}
&&\int dk_1^{\prime -}f(k_1^{\prime -})\langle k_1^{\prime -}k_1^{\prime +}%
\vec{k}_{1\perp }^{\prime }\left| G_0(K)|g_0(K)^{-1}\right| k_1^{+}\vec{k}%
_{1\perp }\rangle =\frac i{2\pi }\delta \left( k_1^{\prime +}-k_1^{+}\right)
\delta \left( \vec{k}_{1\perp }^{\prime }-\vec{k}_{1\perp }\right)  \nonumber
\\
&\times &\int dk_1^{\prime -}\frac{f(k_1^{\prime -})}{\left( k_1^{\prime
-}-k_{1on}^{-}+io\right) }\frac 1{\left( K^{-}-k_1^{\prime
-}-k_{2on}^{-}+io\right) }\left( K^{-}-k_{1on}^{-}-k_{2on}^{-}+io\right) \ .
\label{a2}
\end{eqnarray}
Without any loss of generality, we can think of $f(k_1^{\prime -})$ 
as being split
into a part $f_{uhp}(k_1^{\prime -})$ having singularities only in the upper
half $k_1^{\prime -}$-plane and a part$f_{lhp}(k_1^{\prime -})$ having
singularities only in the lower half $k_1^{\prime -}$-plane, i.e., 
\begin{eqnarray}
f(k_1^{\prime -})=f_{uhp}(k_1^{\prime -})+f_{lhp}(k_1^{\prime -})\ .
\label{a3a}
\end{eqnarray}
In the case that there is a part with poles simultaneously in both half planes,
they can be fully separated, i.e., 
\begin{eqnarray}
&&g(k_1^{\prime -})\frac 1{k_1^{\prime -}-\alpha _1-i\alpha _2}\frac 1{%
k_1^{\prime -}-\beta _1+i\beta _2}=  \nonumber \\
&=&g(k_1^{\prime -})\frac 1{(\alpha -\beta )+i(\alpha _2+\beta _2)}\left[ 
\frac 1{k_1^{\prime -}-\alpha _1-i\alpha _2}-\frac 1{k_1^{\prime -}-\beta
_1+i\beta _2}\right]  \label{a3b}
\end{eqnarray}
with $g(k_1^{\prime -})$ being singularity free. The integration in Eq.(\ref
{a2}) can now be carried out using Cauchy's theorem: 
\begin{eqnarray}
&&\int dk_1^{\prime -}f(k_1^{\prime -})\langle k_1^{\prime -}k_1^{\prime +}%
\vec{k}_{1\perp }^{\prime }\left| G_0(K)|g_0(K)^{-1}\right| k_1^{+}\vec{k}%
_{1\perp }\rangle =\delta \left( k_1^{\prime +}-k_1^{+}\right) \delta \left( 
\vec{k}_{1\perp }^{\prime }-\vec{k}_{1\perp }\right) \times  \nonumber \\
&&\left( K^{-}-k_{1on}^{-}-k_{2on}^{-}+io\right) \times  \nonumber \\
&&\left[ f_{uhp}(k_{1on}^{-})\frac 1{K^{-}-k_{1on}^{-}-k_{2on}^{-}+io}%
+f_{lhp}(K^{-}-k_{2on}^{-})\frac 1{K^{-}-k_{2on}^{-}-k_{1on}^{-}+io}\right] 
\nonumber \\
&=&\delta \left( k_1^{\prime +}-k_1^{+}\right) \delta \left( \vec{k}_{1\perp
}^{\prime }-\vec{k}_{1\perp }\right) \left[
f_{uhp}(k_{1on}^{-})+f_{lhp}(K^{-}-k_{2on}^{-})\right] \ .  \label{a4b}
\end{eqnarray}
We note that propagators cancel and no singularity remains. However, the
result (\ref{a4b}) is for practical purposes useless, since the split into
two parts with disjoint singularities is not known in a numerical
calculation. If, however, the light-front ''energy'' is on-shell, $%
K^{-}=K_{on}^{-}$, then the two terms can be recombined to the original
function, i.e., 
\begin{eqnarray}
\int dk_1^{\prime -}f(k_1^{\prime -})\langle k_1^{\prime -}k_1^{\prime +}%
\vec{k}_{1\perp }^{\prime }\left| G_0(K)|g_0(K)^{-1}\right| k_1^{+}\vec{k}%
_{1\perp }\rangle =\delta \left( k_1^{\prime +}-k_1^{+}\right) \delta \left( 
\vec{k}_{1\perp }^{\prime }-\vec{k}_{1\perp }\right) f(k_{1on}^{-})
\label{a4d}
\end{eqnarray}
for $K^{-}=K_{on}^{-}$.

\newpage

\section{Interaction in first order}

The interaction $w(k)$, defined by Eqs.(\ref{3.4}) and (\ref{2.3a}) to lowest
order of the driving term $V(K)$, is given by 
\begin{eqnarray}
w^{(2)}(K)=g_0(K)^{-1}|G_0(K)V(K)G_0(K)|g_0(K)^{-1}\ ,  \label{b1}
\end{eqnarray}
where the matrix element of the operator $|G_0(K)V(K)G_0(K)|$ is 
\begin{eqnarray}
&&\langle k_1^{\prime +}\vec{k}_{1\perp }^{\prime }\left| \
|G_0(K)V(K)G_0(K)|\ \right| k_1^{+}\vec{k}_{1\perp }\rangle =i\frac{(ig_S)^2%
}{\left( 2\pi \right) ^2}\int dk_1^{\prime -}dk_1^{-}  \nonumber \\
&\times &\frac 1{k_1^{\prime +}(K^{+}-k_1^{\prime +})}\frac 1{\left(
k_1^{\prime -}-\frac{\vec{k}_{1\perp }^{\prime 2}+m_1^2-io}{k_1^{\prime +}}%
\right) }\frac 1{\left( K^{-}-k_1^{\prime -}-\frac{(\vec{K}_{\perp }-\vec{k}%
_{1\perp }^{\prime })^2+m_2^2-io}{K^{+}-k_1^{\prime +}}\right) }  \nonumber
\\
&\times &\frac 1{(k_1^{\prime +}-k_1^{+})}\frac 1{\left( k_1^{\prime
-}-k_1^{-}-\frac{\left( \vec{k}_1^{\prime }-\vec{k}_{1\bot }\right) ^2+\mu
^2-io}{k_1^{\prime +}-k_1^{+}}\right) }  \nonumber \\
&\times &\frac 1{k_1^{+}(K^{+}-k_1^{+})}\frac 1{\left( k_1^{-}-\frac{\vec{k}%
_{1\perp }^2+m_1^2-io}{k_1^{+}}\right) }\frac 1{\left( K^{-}-k_1^{-}-\frac{(%
\vec{K}_{\perp }-\vec{k}_{1\perp })^2+m_2^2-io}{K^{+}-k_1^{+}}\right) }\ .
\label{b2}
\end{eqnarray}

The double integration in $k^-$ in Eq.(\ref{b2}) is performed analytically
using Cauchy's theorem and the condition $K^+ >0$. The integration is
nonzero for $K^+>k^{\prime +}_1>0$ and $K^+>k^+_1>0$. Two possibilities also
appear for $\sigma$ forward propagation. For $k^+_1>k^{\prime +}_1$,
a $\sigma$ is
emitted by particle 1 and otherwise absorbed: 
\begin{eqnarray}
& &\langle k^{\prime +}_1 \vec k^{\prime}_{1\perp} \left| \
|G_0(K)V(K)G_0(K)| \ \right| k^+_1 \vec k_{1\perp} \rangle =\left( ig_S
\right) ^2 \frac{i \theta(K^+- k^{\prime +}_1) \theta( k^{\prime +}_1)} {{k}%
^{\prime +}_{1}(K^+-{k}^{\prime +}_{1}) \left( K^--{k}_{1on}^{\prime -}-{k}%
_{2on}^{\prime -}+io \right) }  \nonumber \\
&\times& \left( \frac{\theta (k^{+}_1-k^{\prime +}_1)}{\left(
k^{+}_1-k^{\prime +}_1 \right) }\frac i{\left( K^{-}-k^{\prime
-}_{1on}-k^{-}_{2on}-k^{\prime -}_{\sigma on} +io\right) }+ \frac{\theta
(k^{\prime +}_1-k^{+}_1)}{\left( k^{\prime +}_1-k^{ +}_1 \right) }\frac i{%
\left( K^{-}-k^{-}_{1on}-k^{\prime -}_{2on}-k^{ -}_{\sigma on}+io\right) }
\right)  \nonumber \\
&\times& \frac{i \theta(K^+-k^+_1) \theta(k^+_1)} {{k}^+_{1}(K^+-{k}^+_{1})
\left(K^--{k}_{1on}^--{k}_{2on}^-+io \right) } \ ,  \label{b3}
\end{eqnarray}
where the light-front "energies" of the intermediate states of the
individual particles are given by 
\begin{eqnarray}
k^{\prime -}_{1on} &=&\frac{\vec k^{\prime 2}_{1\perp}+m_1^2}{k^{\prime +}_1}
\ ,  \nonumber \\
k^{-}_{1on}&=&\frac{\vec k^{2}_{1\perp}+m_1^2}{k^{ +}_1} \ ,  \nonumber \\
k^{\prime -}_{2on}&=&\frac{(\vec K_\perp-\vec k^{\prime }_{1\perp})^2+m_2^2%
} {K^+-k^{\prime +}_1} \ ,  \nonumber \\
k^{-}_{2on}&=&\frac{(\vec K_\perp-\vec k_{1\perp})^2+m_2^2} {K^+-k^{+}_1} \ ,
\nonumber \\
k^{\prime -}_{\sigma on}&=&\frac{ (\vec k^{\prime}_{1\perp} -\vec k_{1\perp}
)^2+\mu^2}{k^+_1-k^{\prime +}_1} \ ,  \nonumber \\
k^{ -}_{\sigma on}&=&\frac{ (\vec k^{\prime}_{1\perp} -\vec k_{1\perp} )^2
+\mu^2}{k^{\prime +}_1-k^{ +}_1} \ .  \label{b4}
\end{eqnarray}
The global three-particle propagator for 1, 2 and $\sigma$ appears in Eq.(%
\ref{b3}), in two cases: when $\sigma$ is either
emitted or absorbed by particle 1.

The matrix element $\langle k^{\prime +}_1 \vec k^{\prime}_{1\perp} \left|
w^{(2)}(K) \right| k^+_1 \vec k_{1\perp} \rangle $ is obtained from Eq.(\ref
{b3}) by multiplying both sides by the matrix element of the operator $%
g_0(K)^{-1}$ from Eq.(\ref{2.11b}). 
\begin{eqnarray}
& &\langle k^{\prime +}_1 \vec k^{\prime}_{1\perp} \left| w^{(2)}(K) \right|
k^+_1 \vec k_{1\perp} \rangle =  \nonumber \\
&=& \left( ig_S\right) ^2 \frac{\theta (k^{+}_1-k^{\prime +}_1)}{\left(
k^{+}_1-k^{\prime +}_1 \right) }\frac i{\left( K^{-}-k^{\prime
-}_{1on}-k^{-}_{2on}-k^{\prime -}_{\sigma on} +io\right) }  \nonumber \\
&+&\left( ig_S\right) ^2 \frac{\theta (k^{\prime +}_1-k^{+}_1)}{\left(
k^{\prime +}_1-k^{ +}_1 \right) }\frac i{\left( K^{-}-k^{-}_{1on}-k^{\prime
-}_{2on}-k^{ -}_{\sigma on}+io\right) }  \nonumber \\
&=& \left( ig_S\right) ^2 \frac{\theta (k^{+}_1-k^{\prime +}_1)}{\left(
k^{+}_1-k^{\prime +}_1 \right) }\frac i{\left( K^{-} -\frac{\vec k^{\prime
2}_{1\perp}+m_1^2}{k^{\prime +}_1} -\frac{(\vec K_\perp-\vec k%
_{1\perp})^2+m_2^2}{K^+-k^{+}_1} -\frac{ (\vec k^{\prime}_{1\perp} -\vec k%
_{1\perp} )^2 +\mu^2}{k^+_1-k^{\prime +}_1} +io\right) }  \nonumber \\
&+&\left( ig_S\right) ^2 \frac{\theta (k^{\prime +}_1-k^{+}_1)}{\left(
k^{\prime +}_1-k^{ +}_1 \right) }\frac i{\left( K^{-} -\frac{\vec k%
^{2}_{1\perp}+m_1^2}{k^{ +}_1} -\frac{(\vec K_\perp-\vec k^{\prime
}_{1\perp})^2+m_2^2}{K^+-k^{\prime +}_1} -\frac{ (\vec k^{\prime}_{1\perp} -%
\vec k_{1\perp} )^2+\mu^2} {k^{\prime +}_1-k^{ +}_1} +io\right) } \ .
\label{b5}
\end{eqnarray}

\newpage

\section{Interaction in second order}

The interaction $w(k)$, defined by Eqs.(\ref{3.4}) and (\ref{2.3a}) to second
order in the driving term $V(K)$, is given by 
\begin{eqnarray}
w(K)\simeq w^{(2)}(K) +w^{(4)}(K)  \label{c1}
\end{eqnarray}
where $w^{(2)}(K)$ is given by Eq.(\ref{b5}) and 
\begin{eqnarray}
w^{(4)}(K)&=&g_0(K)^{-1}|G_0(K)V(K) G_0(K)V(K)G_0(K)|g_0(K)^{-1}  \nonumber
\\
&-&g_0(K)^{-1}|G_0(K)V(K)\widetilde G_0(K)V(K)G_0(K)|g_0(K)^{-1} \ .
\label{c2}
\end{eqnarray}
The second term in Eq.(\ref{c2}) corresponds to the iteration of the
interaction $w^{(2)}(K)$ 
\begin{eqnarray}
& & g_0(K)^{-1}|G_0(K)V(K)\widetilde G_0(K)V(K) G_0(K)|g_0(K)^{-1}= 
\nonumber \\
&=&g_0(K)^{-1}|G_0(K)V(K)G_0(K)|g_0(K)^{-1}|G_0(K)V(K)G_0(K)|g_0(K)^{-1} 
\nonumber \\
&=&w^{(2)}g_0(K)w^{(2)} \ .  \label{c3}
\end{eqnarray}

The matrix element of the operator $|G_0(K)V(K)G_0(K)V(K)G_0(K)|$ is 
\begin{eqnarray}
&&\langle k_1^{\prime +}\vec{k}_{1\perp }^{\prime }\left| \
|G_0(K)V(K)G_0(K)V(K)G_0(K)|\ \right| k_1^{+}\vec{k}_{1\perp }\rangle =\frac{%
(ig_S)^4}{2(2\pi )^6}\int dk_1^{\prime
-}dp_1^{-}dk_1^{-}dp_1^{+}d^2p_{1\perp }  \nonumber \\
&\times &\frac 1{k_1^{\prime +}\left( K^{+}-k_1^{\prime +}\right) }\frac 1{%
\left( k_1^{\prime -}-\frac{\vec{k}_{1\perp }^{\prime 2}+m_1^2-io}{%
k_1^{\prime +}}\right) }\frac 1{\left( K^{-}-k_1^{\prime -}-\frac{(\vec{K}%
_{\perp }-\vec{k}_{1\perp }^{\prime })^2+m_2^2-io}{K^{+}-k_1^{\prime +}}%
\right) }  \nonumber \\
&\times &\frac 1{\left( k_1^{\prime +}-p_1^{+}\right) }\frac 1{\left(
k_1^{\prime -}-p_1^{-}-\frac{(\vec{k}_{1\perp }^{\prime }-\vec{p}_{1\perp
})^2+\mu ^2-io}{k_1^{\prime +}-p_1^{+}}\right) }  \nonumber \\
&\times &\frac 1{p_1^{+}\left( K^{+}-p_1^{+}\right) }\frac 1{\left( p_1^{-}-%
\frac{\vec{p}_{1\perp }^2+m_1^2-io}{p_1^{+}}\right) }\frac 1{\left(
K^{-}-p_1^{-}-\frac{(\vec{K}_{\perp }-\vec{p}_{1\perp })^2+m_2^2-io}{%
K^{+}-p_1^{+}}\right) }  \nonumber \\
&\times &\frac 1{\left( p_1^{+}-k_1^{+}\right) }\frac 1{\left(
p_1^{-}-k_1^{-}-\frac{(\vec{p}_{1\perp }-\vec{k}_{1\perp })^2+\mu ^2-io}{%
p_1^{+}-k_1^{+}}\right) }  \nonumber \\
&\times &\frac 1{k_1^{+}\left( K^{+}-k_1^{+}\right) }\frac 1{\left( k_1^{-}-%
\frac{\vec{k}_{1\perp }^2+m_1^2-io}{k_1^{+}}\right) }\frac 1{\left(
K^{-}-k_1^{-}-\frac{(\vec{K}_{\perp }-\vec{k}_{1\perp })^2+m_2^2-io}{%
K^{+}-k_1^{+}}\right) }\ .  \label{c4}
\end{eqnarray}
The on-energy-shell values of the light-front minus momentum in Eq.(\ref{c4}%
) are given in Eq.(\ref{b4}), and 
\begin{eqnarray}
p_{1on}^{-} &=&\frac{\overrightarrow{p}_{1\bot }^2+m_1^2}{p_1^{+}}\ , 
\nonumber \\
p_{2on}^{-} &=&\frac{(\vec{K}_{\perp }-\vec{p}_{1\perp })^2+m_2^2}{%
K^{+}-p_1^{+}}\ .  \label{cc4}
\end{eqnarray}

The matrix element $\langle k^{\prime +}_1 \vec k^{\prime}_{1\perp} \left| \
|G_0(K)V(K)G_0(K)V(K)G_0(K)| \ \right| k^+_1 \vec k_{1\perp} \rangle$ is
found by analytical integration in the light-front ``energies'' 
in Eq.(\ref{c4}). 
To separate the intermediate four particle propagation, which occurs for $%
k^{\prime +}_1, \ p^{+}_1$ and $k_1^{+}$ satisfying $0<k^{+}_1<p^{+}_1<k^{%
\prime +}<K^{+}$, the following factorization is necessary 
\begin{eqnarray}
& &\frac 1{K^{-}-p^{-}_1 -\frac{\left(\vec K_\perp-\vec p_{1\perp}%
\right)^2+m_2^2-io }{K^+-p^+_1}} \times \frac 1{p^{-}_1-k^-_1 -\frac{\left(%
\vec k_{1\perp}-\vec p_1\right)^2+\mu^2-io}{p_1^+-k_1^{+}}}  \nonumber \\
&=&\frac 1{K^{-}-k^-_1 -\frac{\left( \vec K_\perp-\vec p_{1\perp}%
\right)^2+m_2^2-io }{K^+-p_1^{+}}-\frac{\left(\vec k_{1\perp}-\vec p%
_{1\perp}\right)^2+\mu ^2-io }{ p^+_1-k_1^{+}}}  \nonumber \\
&\times &\left[ \frac 1{K^{-}-p_1^{-} -\frac{\left(\vec K_\perp-\vec p%
_{1\perp}\right)^2+m_2^2-io}{K^+-p_1^{+}}} +\frac 1{p_1^{-}-k_1^--\frac{%
\left(\vec k_{1\perp}-\vec p_{1\perp}\right)^2 +\mu^2-io } {p^+_1-k ^{+}_1}}%
\right] \ .  \label{c5}
\end{eqnarray}

After the Cauchy integration in the light-front ``energies'' the result for $%
\langle k_1^{\prime +}\vec{k}_{1\perp }^{\prime }\left| \
|G_0(K)V(K)G_0(K)V(K)G_0(K)|\ \right| k_1^{+}\vec{k}_{1\perp }\rangle $ in
the region of $0<k_1^{+}<p_1^{+}<k_1^{\prime +}<K^{+}$, which is denoted by $%
\langle k_1^{\prime +}\vec{k}_{1\perp }^{\prime }\left|
|G_0(K)V(K)G_0(K)V(K)G_0(K)|_{(a)}\ \right| k_1^{+}\vec{k}_{1\perp }\rangle $%
, is given by 
\begin{eqnarray}
&&\langle k_1^{\prime +}\vec{k}_{1\perp }^{\prime }\left| \
|G_0(K)V(K)G_0(K)V(K)G_0(K)|_{(a)}\ \right| k_1^{+}\vec{k}_{1\perp }\rangle =
\nonumber \\
&=&\frac{(ig_S)^4}{2(2\pi )^3}\int dp_1^{+}d^2p_{1\perp }\frac{\theta
(k_1^{\prime +})\theta (K^{+}-k_1^{\prime +})}{k_1^{\prime
+}(K^{+}-k_1^{\prime +})}\frac i{K^{-}-\frac{\vec{k}_{1\perp }^{\prime
2}+m_1^2}{k_1^{\prime +}}-\frac{\left( \vec{K}_{\perp }-\vec{k}_{1\perp
}^{\prime }\right) ^2+m_2^2}{K^{+}-k_1^{\prime +}}+io}  \nonumber \\
&\times &\left[ {F}^{\prime }(K)+{F}^{\prime \prime }(K)\right] \frac{\theta
(k_1^{+})\theta (K^{+}-k_1^{+})}{k_1^{+}(K^{+}-k_1^{+})}\frac i{K^{-}-\frac{%
\vec{k}_{1\perp }^2+m_1^2}{k_1^{+}}-\frac{\left( \vec{K}_{\perp }-\vec{k}%
_{1\perp }\right) ^2+m_2^2}{K^{+}-k_1^{+}}+io}\ ,  \label{c6}
\end{eqnarray}
with 
\begin{eqnarray}
{F}^{\prime }(K) &=&\frac{\theta (k_1^{\prime +}-p_1^{+})}{(k_1^{\prime
+}-p_1^{+})}\frac i{K^{-}-\frac{\vec{p}_{1\perp }^2+m_1^2}{p_1^{+}}-\frac{%
\left( \vec{K}_{\perp }-\vec{k}_{1\perp }^{\prime }\right) ^2+m_2^2}{%
K^{+}-k_1^{\prime +}}-\frac{\left( \vec{k}_{1\perp }^{\prime }-\vec{p}%
_{1\perp }\right) ^2+\mu ^2}{k_1^{\prime +}-p_1^{+}}+io}  \nonumber \\
&\times &\frac{\theta (p_1^{+})\theta (K^{+}-p_1^{+})}{p_1^{+}(K^{+}-p_1^{+})%
}\frac i{K^{-}-\frac{\vec{p}_{1\perp }^2+m_1^2}{p_1^{+}}-\frac{\left( \vec{K}%
_{\perp }-\vec{p}_{1\perp }\right) ^2+m_2^2}{K^{+}-p_1^{+}}+io}  \nonumber \\
&\times &\frac{\theta (p_1^{+}-k_1^{+})}{(p_1^{+}-k_1^{+})}\frac i{K^{-}-%
\frac{\vec{k}_{1\perp }^2+m_1^2}{k_1^{+}}-\frac{\left( \vec{K}_{\perp }-\vec{%
p}_{1\perp }\right) ^2+m_2^2}{K^{+}-p_1^{+}}-\frac{\left( \vec{p}_{1\perp }-%
\vec{k}_{1\perp }\right) ^2+\mu ^2}{p_1^{+}-k_1^{+}}+io}\ ;  \label{c7}
\end{eqnarray}
\begin{eqnarray}
{F}^{\prime \prime }(K) &=&\frac{\theta (k_1^{\prime +}-p_1^{+})}{%
(k_1^{\prime +}-p_1^{+})}\frac i{K^{-}-\frac{\vec{p}_{1\perp }^2+m_1^2}{%
p_1^{+}}-\frac{\left( \vec{K}_{\perp }-\vec{k}_{1\perp }^{\prime }\right)
^2+m_2^2}{K^{+}-k_1^{\prime +}}-\frac{\left( \vec{k}_{1\perp }^{\prime }-%
\vec{p}_{1\perp }\right) ^2+\mu ^2}{k_1^{\prime +}-p_1^{+}}+io}  \nonumber \\
&\times &\frac i{K^{-}-\frac{\vec{k}_{1\perp }^2+m_1^2}{k_1^{+}}-\frac{%
\left( \vec{K}_{\perp }-\vec{k}_{1\perp }^{\prime }\right) ^2+m_2^2}{%
K^{+}-k_1^{\prime +}}-\frac{\left( \vec{k}_{1\perp }^{\prime }-\vec{p}%
_{1\perp }\right) ^2+\mu ^2}{k_1^{\prime +}-p_1^{+}}-\frac{\left( \vec{p}%
_{1\perp }-\vec{k}_{1\perp }\right) ^2+\mu ^2}{p_1^{+}-k_1^{+}}+io} 
\nonumber \\
&\times &\frac{\theta (p_1^{+}-k_1^{+})}{(p_1^{+}-k_1^{+})}\frac i{K^{-}-%
\frac{\vec{k}_{1\perp }^2+m_1^2}{k_1^{+}}-\frac{\left( \vec{K}_{\perp }-\vec{%
p}_{1\perp }\right) ^2+m_2^2}{K^{+}-p_1^{+}}-\frac{\left( \vec{p}_{1\perp }-%
\vec{k}_{1\perp }\right) ^2+\mu ^2}{p_1^{+}-k_1^{+}}+io}  \label{c8}
\end{eqnarray}

The part of the propagator given by Eq.(\ref{c6}) contains the virtual
light-front propagation of intermediate states with up to four particles.
The function $F^{\prime}$ contains only intermediate states up to three
particles and is two-body reducible. It  will eventually be canceled by the
corresponding piece in the second term in (\ref{c2}). The function $%
F^{\prime\prime}$ has one intermediate state in which
four-particle propagator which
can be recognized as the middle piece of Eq.(\ref{c8}). The other
possibility which includes up to four particles in the intermediate state
propagation is given by $0<k^{\prime+}_1<p_1^{+}<k_1^{+}<K^{+}$. To obtain
this part, we perform the 
transformation $k^\prime_1\leftrightarrow k_1$ in Eq.(\ref{c6}).

The contribution of the region $0<p^{+}_1<k_1^{+}<k^{\prime +}_1<K^{+}$ to
the matrix element $\langle k^{\prime +}_1 \vec k^{\prime}_{1\perp} \left| \
|G_0(K)V(K)G_0(K)V(K)G_0(K)| \ \right| k^+_1 \vec k_{1\perp} \rangle $ is
denoted by $\langle k^{\prime +}_1 \vec k^{\prime}_{1\perp} \left|
|G_0(K)V(K)G_0(K)V(K)G_0(K)|_{(b)} \ \right| k^+_1 \vec k_{1\perp} \rangle $.
It contains only up to three-particle intermediate states and is two-body
reducible. Consequently, it will be canceled by the corresponding piece of the
second term in (\ref{c2}). It is given by 
\begin{eqnarray}
& &\langle k^{\prime +}_1 \vec k^{\prime}_{1\perp} \left|
|G_0(K)V(K)G_0(K)V(K)G_0(K)|_{(b)} \ \right| k^+_1 \vec k_{1\perp} \rangle =
\nonumber \\
&=&\frac{(ig_S)^4}{2(2\pi)^3}\int dp^+_1d^2p_{1\perp} \frac {\theta
(k_1^{\prime +})\theta (K^{+}-k_1^{\prime +})} {k^{\prime
+}_1(K^+-k_1^{\prime +})}\frac{i} {K^{-}-\frac{\vec k_{1\perp }^{\prime
2}+m_1^2 }{k_1^{\prime +}} -\frac{\left(\vec K_\perp-k^{\prime
}_{1\perp}\right)^2+m_2^2 } { K^+-k_1^{\prime +}}+io}  \nonumber \\
&\times &\frac{\theta (k^{\prime +}_1-k_1^{+})}{(k^{\prime +}_1-k_1^{+})} 
\frac i{K^{-}-\frac{\vec p_{1\perp }^2+m_1^2}{p^{+}_1} -\frac{\left(\vec K%
_\perp-\vec k^{\prime}_{1\perp}\right)^2+m_2^2} { K^+-k^{\prime +}_1} -\frac{%
\left(\vec k^\prime_{1\perp}-\vec p_{1\perp}\right)^2 +\mu^2}{k^{\prime
+}_1-p_1^{+}}+io}  \nonumber \\
&\times & \frac{\theta(p^+_1)\theta(K^+-p^+_1)}{p^+_1(K^+-p^+_1)} \frac i{%
K^{-}-\frac{\vec p_{1\perp }^2+m_1^2}{p_1^{+}} -\frac{\left(\vec K_\perp-%
\vec p_{1\perp}\right)^2+m_2^2}{ K^+-p_1^{+}}+io}  \nonumber \\
&\times &\frac{\theta(k^+_1-p_1^{+})}{(k^+_1-p_1^{+})} \frac i{K^{-}-\frac{%
\vec p_{1\perp }^2+m_1^2}{p_1^{+}} -\frac{\left( \vec K_\perp-\vec k%
_{1\perp}\right)^2+m_2^2} { K^+-k^{+}_1}-\frac{\left(\vec k_{1\perp}-\vec p%
_{1\perp}\right)^2 +\mu^2}{k^+_1-p_1^{+}}+io}  \nonumber \\
&\times & \frac{\theta (k^+_1)\theta (K^{+}-k_1^{+})}{k_1^{+}(K^+-k_1^+)}%
\frac{i}{K^{-}-\frac{\vec k_{1\perp}^2+m_1^2 } {k^{+}_1}-\frac{\left( \vec K%
_\perp-\vec k_{1\perp}\right)^2+m_2^2 } { K^+-k_1^+ }} \ .  \label{c9}
\end{eqnarray}

For the momentum region satisfying $0<k^{\prime +}_1<k^{+}_1<p_1^{+}<K^{+}$,
the contribution to the matrix element $\langle k^{\prime +}_1 \vec k%
^{\prime}_{1\perp} \left| \ |G_0(K)V(K)G_0(K)V(K)G_0(K)| \ \right| k^+_1 
\vec k_{1\perp} \rangle $ can be obtained from Eq.(\ref{c9}) by performing the
following
transformation on the kinematical momentum: $k^{\prime}_1\leftrightarrow
K-k^{\prime}_1$, $k_1\leftrightarrow K-k_1$ and $m_1\leftrightarrow m_2$.
From Eqs. (\ref{c8}) and (\ref{c9}), the following result is obtained 
\begin{eqnarray}
& &\langle k^{\prime +}_1 \vec k^{\prime}_{1\perp} \left| \
|G_0(K)V(K)G_0(K)V(K)G_0(K)| \ \right| k^+_1 \vec k_{1\perp} \rangle = 
\nonumber \\
&=& \left( \langle k^{\prime +}_1 \vec k^{\prime}_{1\perp} \left| \
|G_0(K)V(K)G_0(K)V(K)G_0(K)|_{(a)} \ \right| k^+_1 \vec k_{1\perp} \rangle +
\left[ k^\prime_1 \leftrightarrow k_1 \right] \right)  \nonumber \\
&+& (\langle k^{\prime +}_1 \vec k^{\prime}_{1\perp} \left| \
G_0(K)V(K)G_0(K)V(K)G_0(K)|_{(b)} \ \right| k^+_1 \vec k_{1\perp} \rangle 
\nonumber \\
&+&\left[k^\prime_1\leftrightarrow K-k^\prime_1, \ k_1\leftrightarrow K-k_1
, m_1\leftrightarrow m_2\right] ) \ .  \label{c10}
\end{eqnarray}

The subtraction of the iterated first order driving term in Eq.(\ref{c2})
cancels the corresponding terms in Eq.(\ref{c10}) such that the matrix
element $\langle k^{\prime +}_1 \vec k^{\prime}_{1\perp} \left| w^{(4)}(K)
\right| k^+_1 \vec k_{1\perp} \rangle $ is two-body irreducible with a
global four-body propagation. It is obtained from Eqs.(\ref{c6}), (\ref{c8})
and (\ref{c2}) as
\begin{eqnarray}
& &\langle k^{\prime +}_1 \vec k^{\prime}_{1\perp} \left| w^{(4)}(K) \right|
k^+_1 \vec k_{1\perp} \rangle =  \nonumber \\
&=&\frac{(ig_S)^4}{2(2\pi)^3}\int dp^+_1d^2p_{1\perp} \frac{\theta(k^{\prime
+}_1-p^+_1)}{(k^{\prime +}_1-p^+_1)} \frac i{K^{-}-\frac{\vec p_{1\perp
}^2+m_1^2}{p_1^{+}}- \frac{\left(\vec K_\perp-\vec k^{\prime}_{1\perp}%
\right)^2+m_2^2} {K^+-k_1^{\prime +}}-\frac{\left(\vec k^{\prime}_{1\perp} -%
\vec p_{1\perp}\right)^2+\mu^2}{k^{\prime +}_1-p_1^{+}}+io}  \nonumber \\
&\times & \frac i{K^{-}-\frac{\vec k_{1\perp }^2+m_1^2}{k_1^{+}} -\frac{%
\left(\vec K_\perp-\vec k^{\prime}_{1\perp}\right)^2+m_2^2} { K^+-k^{\prime
+}_1} -\frac{\left(\vec k^{\prime}_{1\perp}-\vec p_{1\perp}\right)^2+\mu ^2%
} { k^{\prime +}_1-p_1^{+}}-\frac{\left(\vec p_{1\perp}-\vec k%
_{1\perp}\right)^2 +\mu^2}{ p_1^+-k_1^{+}}+io}  \nonumber \\
&\times & \frac{\theta(p^+_1-k^+_1)}{(p^+_1-k^+_1)} \frac i{K^{-}-\frac{\vec 
k_{1\perp }^2+m_1^2}{k^{+}_1}-\frac{\left(\vec K_\perp-\vec p%
_{1\perp}\right)^2+m_2^2}{K^+-p^{+}_1}-\frac{\left(\vec p_{1\perp}-\vec k%
_{1\perp}\right)^2+\mu^2}{p^+_1-k_1^{+}}+io}  \nonumber \\
&+&\left[ k^\prime_1 \leftrightarrow k_1 \right] \ .  \label{c11}
\end{eqnarray}

\newpage

\section{Integral Equation for the Bound-State}

In the approximation considered, the vertex function satisfies  an
integral equation with the kernel containing two parts, one corresponding to
Eq.(\ref{b5}) and the other to Eq.(\ref{c11}). The plus momentum are
rescaled by $K^+$, such that the momentum fractions $x=\frac{k^+_1}{K^+}$, $%
y=\frac{k^{\prime +}_1}{K^+}$, and $z=\frac{p^+_1}{K^+}$, are used. The
notation $\left \langle k^{\prime +}_1 \vec k^\prime_{1\perp}\right| 
\gamma _B\rangle\equiv \widetilde\gamma _B(y,\vec k%
^\prime_{1\perp})$ is introduced. The homogeneous integral equation for the
light-front vertex function is evaluated in the center of mass system, 
\begin{equation}
\gamma _B(y,\vec k^\prime_{1\perp}) = \frac{1}{(2\pi)^3}\int \frac{%
d^2k_{1\perp }dx}{2x(1-x)}\frac{{\cal K}^{(2)} (y,\vec k^{\prime}_{1\perp
};x,\vec k_{1\perp })+ {\cal K}^{(4)}(y,\vec k^{\prime}_{1\perp };x,\vec k%
_{1\perp })} {M_B^2-M_0^2}\gamma _B(x,\vec k_{1\perp}) \ ,
\label{bsnp}
\end{equation}
where the free two-body mass is $M^2_0=\frac{\vec k_{1\perp}^2+m^2}{x(1-x)}$
and $0\ < \ x \ < \ 1$.

The part of the kernel which has only the propagation of virtual three
particles states foward in the light-front time is obtained from Eq.(\ref{b5}%
), 
\begin{eqnarray}
&& {\cal K}^{(2)} (y,\vec k^{\prime}_{1\perp };x,\vec k_{1\perp })= 
\nonumber \\
&&g_S^2 \frac{\theta (x-y)}{\left( x-y\right) \left( M_B^2- \frac{\vec k%
^{\prime 2}_{1\perp}+m^2}y-\frac{\vec k_{1\perp }^2+m^2}{1-x}- \frac{(\vec k%
^\prime_{1\perp }-\vec k_{1\perp })^2+\mu ^2}{x-y}\right) } +\left[
x\leftrightarrow y,\vec k^\prime_{1\perp } \leftrightarrow \vec k%
_{1\perp}\right].  \label{k2}
\end{eqnarray}
Eq.(\ref{bsnp}) with the effective interaction given by (\ref{k2})
corresponds to the Weinberg equation derived from the BSE in the infinitum
momentum frame \cite{wein}. It has also been solved in Ref.\cite{ji86} and
in Ref.\cite{ji94} including self-energy correction. The equivalent equation
for fermions has been discussed in Ref.\cite{pe90}. 

The contribution to the kernel from the virtual four-body propagation is
obtained from Eq.(\ref{c11}), 
\begin{eqnarray}
&&{\cal K}^{(4)}(y,\vec k^{\prime}_{1\perp };x,\vec k_{1\perp })=  \nonumber
\\
&&\frac{g_S^4}{(2\pi)^3}\int \frac{d^2p_{\bot }dz}{2z(1-z)\left( z-x\right)
(y-z)}\frac{\theta (z-y)\theta (x-z)}{\left( M_B^2-\frac{\vec k^{\prime
2}_{1\perp }+m^2}y-\frac{\vec p_{1\perp }^2+m^2}{1-z} -\frac{(\vec k%
^\prime_{1\perp }- \vec p_{1\perp })^2+\mu ^2}{z-y}\right) }  \nonumber \\
&\times& \frac 1{\left( M_B^2- \frac{\vec k^{\prime 2}_{1\perp }+m^2}y-\frac{%
\vec k_{1\perp }^2+m^2}{1-x}-\frac{(\vec k^\prime_{1\perp }-\vec p_{1\perp
})^2+\mu ^2}{z-y} -\frac{(\vec p_{1\perp }-\vec k_{1\perp })^2+\mu ^2}{x-z}%
\right) }  \nonumber \\
&\times& \frac 1{\left( M_B^2-\frac{\vec p_{1\perp }^2+m^2}z -\frac{\vec k%
_{1\perp }^2+m^2}{1-x}-\frac{(\vec p_{1\perp }-\vec k_{1\perp })^2+\mu ^2}{%
x-z}\right) }+\left[ x\leftrightarrow y,\vec k_{1\perp }\leftrightarrow \vec 
k^\prime_{1\perp }\right] \ .  \label{k4}
\end{eqnarray}
Eqs.(\ref{bsnp})-(\ref{k4}) are easily recognized to be covariant under
kinematical light-front boosts. However, the covariance of the
four-dimensional wave-function (\ref{3.6b}) is certainly lost by a finite
expansion of $W(K)$ in Eq.(\ref{2.3a}) and the use of the 
 corresponding $w(K)$ while
covariance continues to hold for the solution of Eq.(\ref{2.3a}).

\newpage

\begin{center}
{\bf FIGURE CAPTIONS}
\end{center}

{\bf Fig.1.} Light-front time ordered diagrams for  $w^{(2)}(K)$ (a) and 
$w^{(4)}(K)$ (b), representing the  
light-front time ordered view of one and two $\sigma$ exchanges, 
respectively.

{\bf Fig.2.} Results for $g_S$ as a function of the two-body 
bound state mass $M_B$ for $%
\mu=0.5m$. Numerical solution of the covariant four-dimensional BSE (%
\ref{bs}) (solid curve), the light-front Eq.(\ref{3.5})
with interaction including up to three-particles in the intermediate states, 
i.e., with $w(K_B)\simeq w^{(2)}(K_B)$ (dashed curve) and 
including up to four-particles in the
intermediate states, i.e., with  $w(K_B)\simeq w^{(2)}(K_B)+w^{(4)}(K_B)$ 
(dotted curve). 
Solution of the quantum mechanics squared mass eigenvalue equation
(\ref{5.4}), with $w(K_v)\simeq w^{(2)}(K_v)$ 
(long-dashed curve), and with 
$w(K_v)\simeq w^{(2)}(K_v)+w^{(4)}(K_v)$ 
(short-dashed curve) defining the
two-particle potential in Eq.(\ref{5.5a}).

{\bf Fig.3.}  
Results for the transverse momentum distribution $f(q)$ as a function 
of the transverse component, $q$, of the individual four-momentum, 
for $M_B=0$ and $\mu=0.5m$: (a) numerical solution of the 
four-dimensional BSE  with $g_s=20.14$; (b) relative error of the various
approximations with respect to the four-dimensional BSE results, defined
by $Df(q)=1-f^{(n)}_{app}(q)/f_{exact}(q)$ with $n=2$ and 4. 
Results for the light-front Eq.(\ref{3.5})
with an interaction including up to three-particles 
in the intermediate states, 
i.e., with $w(K_B)\simeq w^{(2)}(K_B)$ where $g_s=20.8$ (dashed curve) and 
with an interaction including up to four-particles in the
intermediate states, i.e., with  $w(K_B)\simeq w^{(2)}(K_B)+w^{(4)}(K_B)$ 
where $g_s=20.2$ (dotted curve). 
Solutions of the quantum mechanics squared mass eigenvalue equation
(\ref{5.4}), with the two-particle potential in Eq.(\ref{5.5a})
defined by $w(K_v)\simeq w^{(2)}(K_v)$ where $g_s=15.7$ 
(long-dashed curve), and with 
$w(K_v)\simeq w^{(2)}(K_v)+w^{(4)}(K_v)$ where $g_s=14.9$
(short-dashed curve).

{\bf Fig.4.} Results for the transverse momentum
distribution $f(q)$ as a function the
of transverse component, $q$, of the individual four-momentum, 
for $M_B=1.98m$ and $\mu=0.5m$: (a) numerical solution of the 
four-dimensional BSE with $g_s=9.03$ ; (b) relative error of the various
approximations in respect to the four-dimensional BSE results, defined
by $Df(q)=1-f^{(n)}_{app}(q)/f_{exact}(q)$ with $n=2$ and 4. 
Results for the light-front Eq.(\ref{3.5})
with interaction including up to three-particles in the intermediate states, 
i.e., with $w(K_B)\simeq w^{(2)}(K_B)$ where $g_s=9.10$ (dashed curve) and 
with an interaction
including up to four-particles in the
intermediate states, i.e., with  $w(K_B)\simeq w^{(2)}(K_B)+w^{(4)}(K_B)$ 
where $g_s=9.03$ (dotted curve). 
Solutions of the quantum mechanics squared mass eigenvalue equation
(\ref{5.4}), with the
two-particle potential in Eq.(\ref{5.5a}) defined by
 $w(K_v)\simeq w^{(2)}(K_v)$ where $g_s=8.33$ 
(long-dashed curve), and with 
$w(K_v)\simeq w^{(2)}(K_v)+w^{(4)}(K_v)$ where $g_s=8.23$
(short-dashed curve). 


\begin{figure}[h]
\centerline{\ 
\begin{picture}(330,200)(0,0)
\Line(90,10)(230,10)
\DashLine(130,10)(190,70){5}
\Line(90,70)(230,70)
\end{picture} 
}
\end{figure}
\begin{center}
{\bf Fig.1a}
\end{center}
\vskip 2cm
\begin{figure}[h]
\centerline{\ 
\begin{picture}(330,130)(0,0)
\Line(90,10)(230,10)
\DashLine(120,10)(180,70){5}
\DashLine(140,10)(200,70){5}
\Line(90,70)(230,70)
\end{picture}
}
\end{figure}
\begin{center}
{\bf Fig.1b}
\end{center}
\postscript{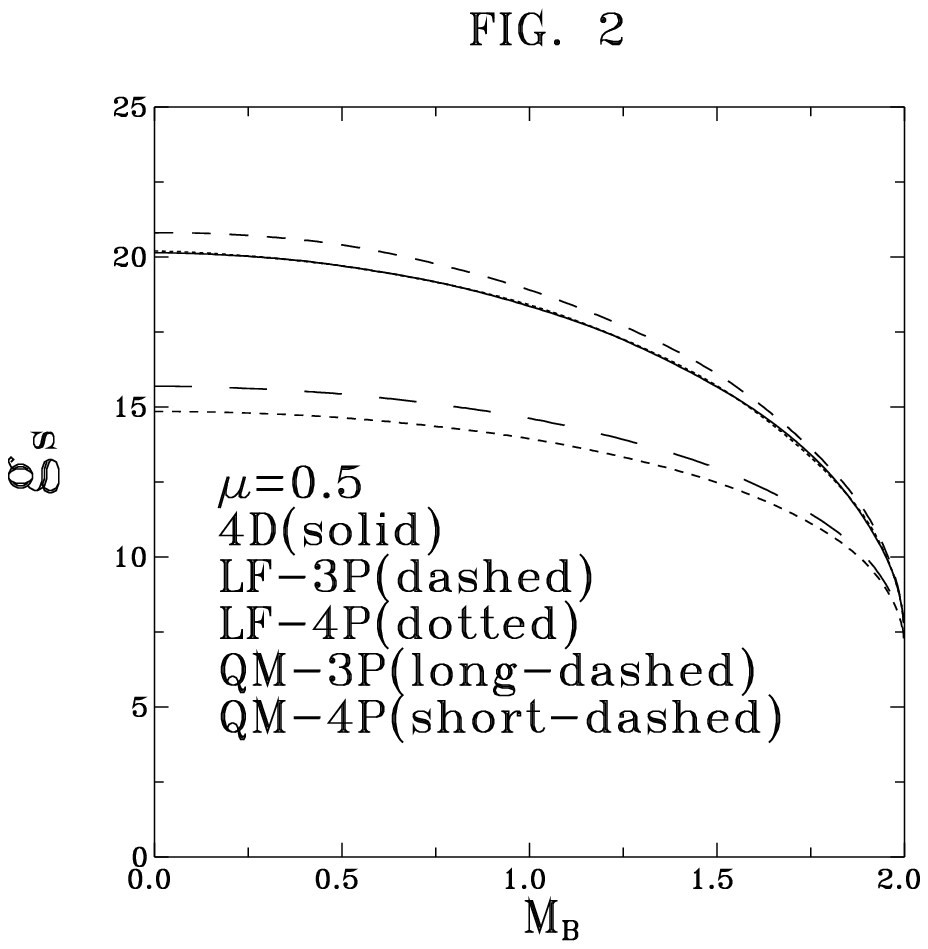}{1.0}    
\vspace{1 true cm} 
\postscript{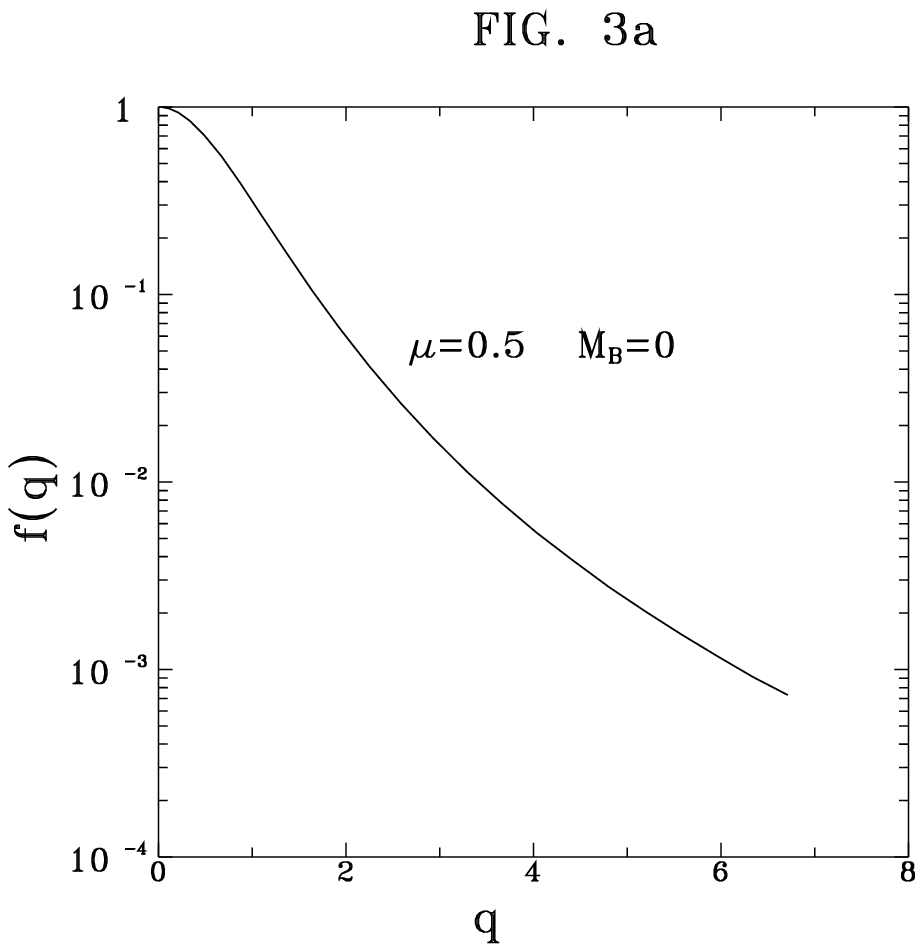}{1.0}    
\vspace{1 true cm} 
\postscript{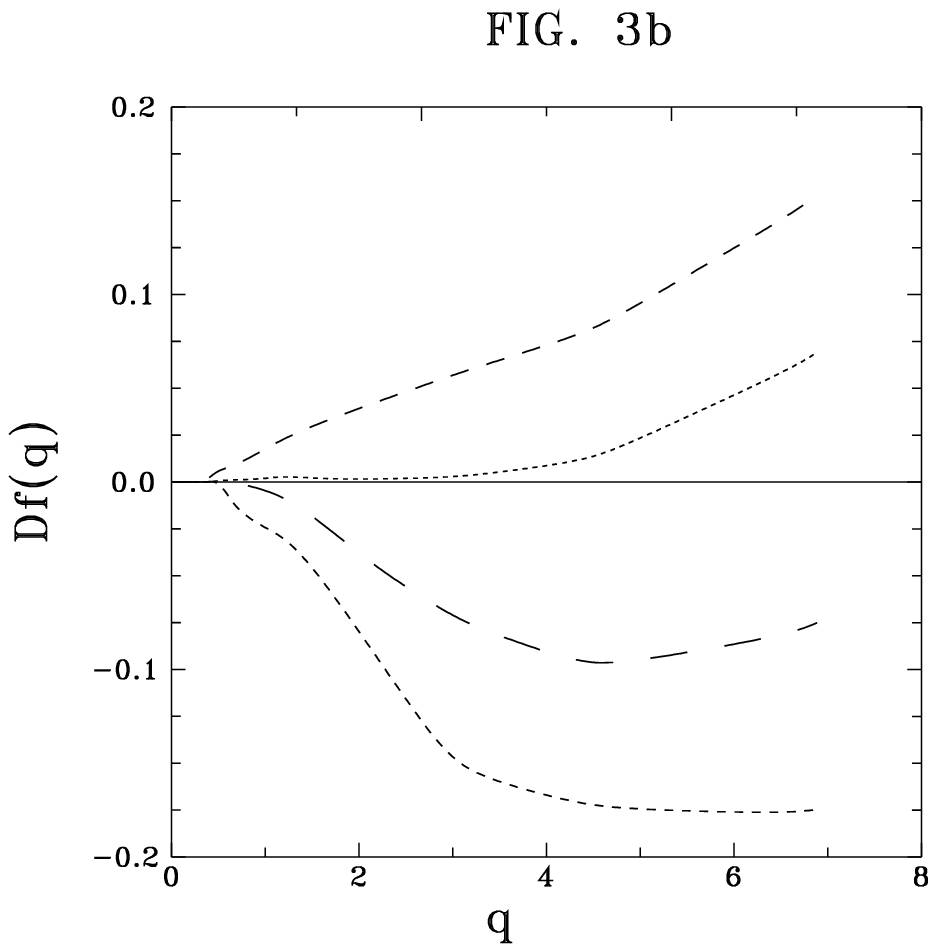}{1.0}    
\vspace{1 true cm} 
\postscript{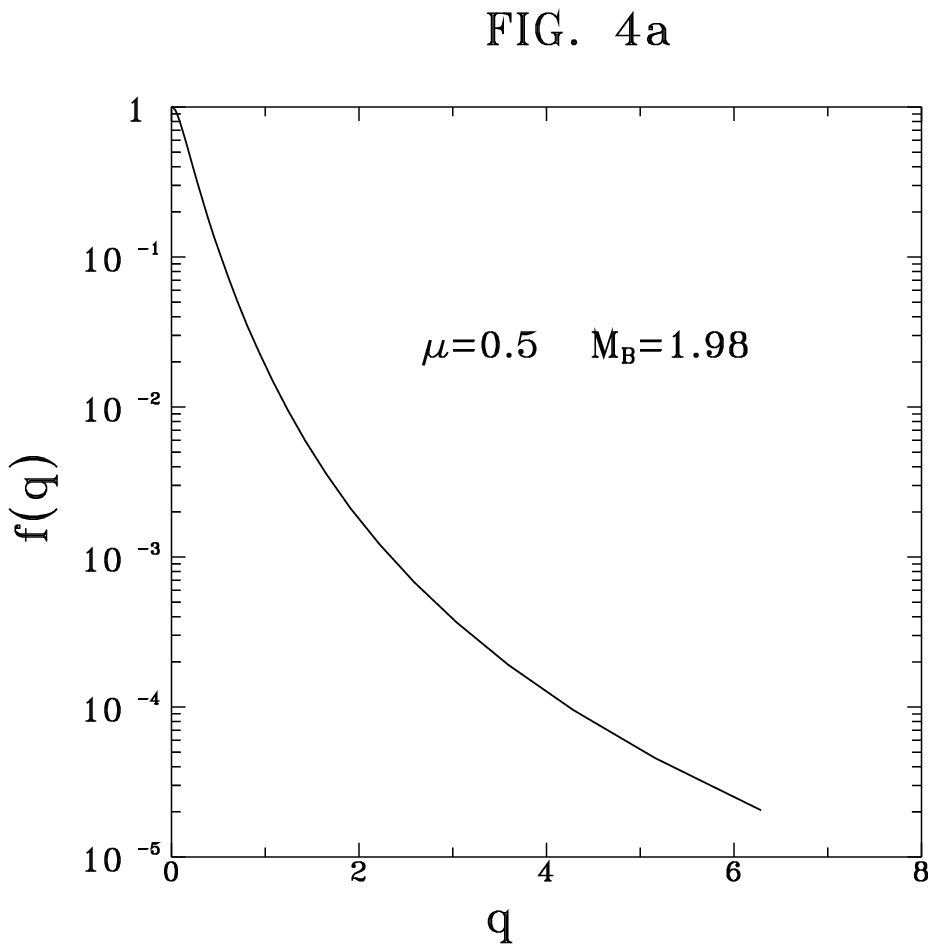}{1.0}    
\vspace{1 true cm} 
\postscript{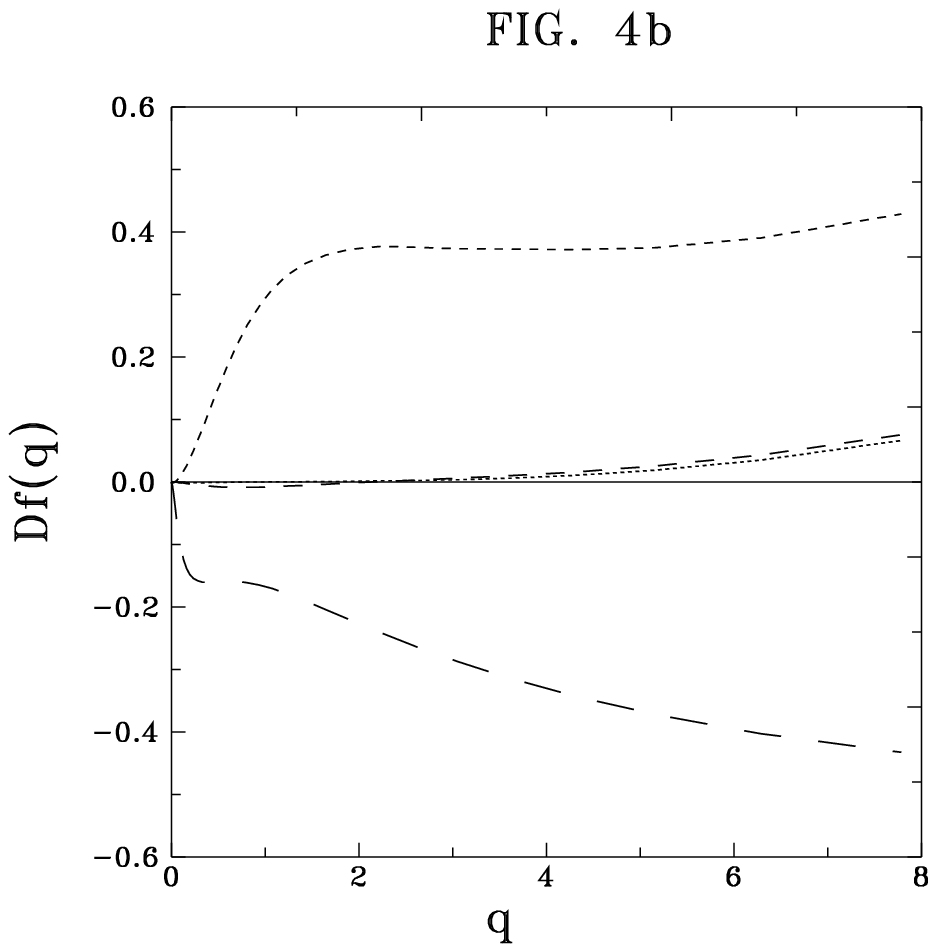}{1.0}    

\end{document}